\documentclass[11pt,a4paper,oneside]{article}
\usepackage{epsfig}
\usepackage{amsfonts}
\usepackage[latin1]{inputenc}
\usepackage{amsmath}
\usepackage{amssymb}

\newtheorem{prop}{\underline{Proposition}}[section]

\newcommand{\R}{\mathbb{R}}
\newcommand{\pr}{\mathbb{P}}

\newcommand{\be}{\begin{equation}}
\newcommand{\ee}{\end{equation}}

\date{}

\title{From Minority Game to Black\&Scholes pricing}
\author{Matteo Ortisi\footnote{Pioneer Investments, Galleria San Carlo 6, 20122 Milano, Italy (matteo.ortisi@pioneerinvestments.com)}, Valerio Zuccolo\footnote{Intesa Sanpaolo, Piazza Paolo Ferrari 10, 20100 Milano, Italy (valerio.zuccolo@polimi.it)}}

\begin{document}
\maketitle
\begin{abstract}
In this paper we study the continuum time dynamics of a stock in a market where agents behavior is modeled by a Minority Game and a Grand Canonical Minority Game. The dynamics derived is a generalized geometric Brownian motion; from the Black\&Scholes formula the calibration of both the Minority Game and the Grand Canonical Minority Game, by means of their characteristic parameters, is performed. 
We conclude that for both games the asymmetric phase with characteristic parameters close to critical ones is coherent with options implied volatility market.

\end{abstract}

\section{Introduction}
In recent years researchers endeavored to build models that reproduce some empirical statistical regularities of the real financial markets, such as volatility clusters, fat tails, scaling, occurrence of crashes, etc., known as ``stylized facts" (see \cite{giardina, john, lux}). One possible way to address this problem is the ``black-box" stochastic approach. A complex stochastic process which possesses the relevant characteristics of the desired
empirical facts is constructed. Processes used in this kind of models are nonlinear diffusion models, L\'evy processes, jump processes and stochastic volatility models. Another possibility is to follow the agent-based framework, which models the market and derives from the interaction between players the stock price dynamics. Herd behavior and Minority Game are examples of these models. Choosing between these two models families relies on the purpose of the modeling itself. ``Black-box" stochastic modeling is used in financial mathematics, where the main goal is practical use, i.e. the pricing of derivatives or portfolio allocation. Agent-based models are mainly used in economics where the theoretical aspects, the explanation and understanding of financial markets, are the main intent.

There is a complementarity between these two approaches; indeed ``black-box" stochastic models lack of explanatory power by construction and agent-based models can be hardly used for pricing purposes given their complexity.
Moreover one of most attractive features of the usual Black\&Scholes type models is the possibility of obtaining closed, exact formulas for the pricing of financial derivative securities. Such qualities are fundamental from the point of view of finance practitioners, like financial institutions, that require fast calculations, closed pricing formula for calibration purposes and tools to hedge risky assets.

A very promising way to combine these main issues, is to focus on highly simplified toy models of financial markets relying on the Minority Game (MG) \cite{challet}. Variants of this model, like the GCMG, have been shown to reproduce quite accurately such stylized facts of financial markets (see \cite{Bouchaud, challet_quan, challet1, Jef, stan}) and  moreover the continuum time limit of the MG and the GCMG provides a bridge between the adherence to empirically observed features of real markets and the practical usability of ``black-box" stochastic models, since its evolution follows a system of stochastic differential equations (see \cite{marsili1}, \cite{coolen}).

In this framework, the stock price process is driven by the excess demand or overall market bid. The excess demand evolution process for real and ``fake'' market histories has been explicitly obtained in \cite{coolen}. While a game with real market histories can fit stock prices better than a game with ``fake'' ones, the latter is more mathematically tractable, especially if the purpose is to obtain closed formulas and not run the game through simulations. Models with ``fake'' market histories, where at each point in time all agents are given random rather than real market data upon which to base their decisions, have the advantage of being Markovian and hence suitable to the application of classic results, like Girsanov theorem, that are at the hearth of option pricing techniques in mathematical finance. On the other side, models with real market histories are strongly non-Markovian and a dynamics simple like (\ref{stoch_eq1.1}) is no more available. The dependence of the process on the history makes meaningless all the averages present in equation (\ref{stoch_eq1.1}); the evolution process becomes a system of stochastic differential equations whose parameters are to be obtained as a solution of a dynamic system defining the bid evolution process (see \cite{coolen} for details). Since our aim is to calibrate the MG and the GCMG on the real market of European call options, as a first attempt to obtain a closed formula for pricing an European call option on a single stock it is easier, and maybe compulsory, to use the simplest version of these games with ``fake'' market histories and to recover the excess demand directly from the scores difference stochastic differential equations obtained in \cite{marsili_GCMG,marsili1}. Using this approach it is possible to apply the usual ``change of measure technique'', that guarantee the absence of arbitrage in the model and the existence of the hedging portfolio, and hence to obtain closed pricing formulas allowing the model calibration to the market. Undoubtedly the MG with real market history must be investigated and the approach used in \cite{coolen} must be kept in mind for further development.

The work is organized as follow. In Section \ref{section_the_model} the classical MG and GCMG models and their continuum time limit versions are described. In Section \ref{stock} we derive the stock price dynamics and in Section \ref{risk-neutral}, by using the classical risk-neutral pricing techniques (see \cite{shreve}), we apply the Black\&Scholes formula to price European options. Section \ref{section_5} contains the process to calibrate the MG and the GCMG on the European options market for several sets of game and option parameters.

\section{The Models}
\label{section_the_model}
In this section the basic concept and the main results of MG and GCMG useful for our purposes will be exposed (for a comprehensive introduction to MG and GCMG refer to \cite{coolenlibro,marsililibro}).
\subsection{Minority Game}
Here we are considering a slightly modified version of the MG, where there are $N$ agents and its dynamics is defined in terms of dynamical variables $U_{s,i}(t)$, scores corresponding to each of the possible agents strategy choices $s=+w,-w$. This version of the game with non-unitary weights has already been studied in \cite{challet_quan}.

Each agent takes a decision (strategy choice) $s_i(t)$ with
$$
Prob\{s_i(t)=s\}=\frac{e^{\Gamma_i U_{s,i}(t)}}{\sum_{s'}e^{\Gamma_iU_{s',i}(t)}}
$$
where $\Gamma_i>0$, and $s'\in \{-w,+w\}$. The original MG corresponds to $\Gamma_i=\infty$ and was generalized to $\Gamma_i=\Gamma<\infty$ \cite{cavagna}; here we consider the latter case. 

The public information variable $\mu(t)$, that represent the common knowledge of the past record, is given to all agents; it belongs to the set of integers $\{1,\ldots,P\}$ and can either be the binary encoding of last $M$ winning choices or drawn at random from a uniform distribution:
$$
Prob\{\mu(t)=\mu\}=\frac{1}{P}, \quad \mu=1,\ldots,P.
$$
Here we consider the latter case (see \cite{cavagna}).

The strategies $a_{s,i}^{\mu}$ are uniform random variables taking values $\pm w$ ($Prob\{a_{s,i}^{\mu}=\pm w \}=1/2$) independent on $i, s$ and $\mu$.
Here we consider $S=2$, i.e. 2 strategy for each agent that are randomly drawn at the beginning of the game and kept fixed.

On the basis of the outcome $B(t)=\sum_{i=1}^Na_{s_i(t),i}^{\mu(t)}$ each agent update his scores according to
\be
\label{increment}
U_{s,i}(t+\delta t)=U_{s,i}(t)-a_{s,i}^{\mu(t)}\frac{B(t)}{P},
\ee
where $\delta t \to 0$ is the time increment.

Let us introduce the following random variables (to ease the notation the choices $+w$ and $-w$ are shorted with $+$ and $-$)
$$\xi_i^{\mu}=\frac{a_{+,i}^{\mu}-a_{-,i}^{\mu}}{2}, \quad \Theta^{\mu}=\sum_{i=1}^N\frac{a_{+,i}^{\mu}+a_{-,i}^{\mu}}{2}$$
and their averages
$$
\overline{\xi_i\Theta}=\frac{1}{P}\sum_{\mu=1}^P\xi_i^{\mu}\Theta^{\mu}, \quad \overline{\xi_i\xi_j}=\frac{1}{P}\sum_{\mu=1}^P\xi_i^{\mu}\xi_j^{\mu}.
$$
The only relevant quantity in the dynamics is the difference between the scores of the two strategies:
\be
\label{diff}
y_i(t)=\Gamma\frac{U_{+,i}(\tau)-U_{-,i}(\tau)}{2},
\ee
where $\tau=\frac{t}{\Gamma}$.

Let $(\Omega,\mathcal{F},\pr)$ be the probability space respect to which all our random variables are defined, $y=(y_i)_{1\leq i\leq N}$, $\Theta=(\Theta^{\mu})_{1\leq\mu\leq P}$ and $\xi=(\xi_i)_{1\leq i\leq N}$.

As shown in \cite{marsili1}, if $P/N=\alpha\in\R_+$,  $S=2$ and $\Gamma_i=\Gamma>0$ for all $i$, the dynamics of the continuum time limit of the MG is given by the following $N$-dimensional stochastic differential equation
\begin{equation}
\label{stoch_eq1.1}
dy_i(t)=\left(-\overline{\xi_i\Theta}-\sum_{j=1}^N\overline{\xi_i\xi_j}\tanh (y_j(t))\right)dt + A_i(\alpha,N,\Gamma,\xi)d{\mathbf W}(t), \quad i=1,\ldots,N
\end{equation}
where
\begin{itemize}
\item[]${\mathbf W}(t)$ is an $N$-dimensional Wiener process,
\item[]$A_i$ is the $i$-th row of the $N\times N$ matrix $A=(A_{ij})$ such that $$(AA')_{ij}(\alpha,N,\Gamma,\xi)=\frac{\Gamma\sigma_{N}^2w^2}{\alpha N}\overline{\xi_i\xi_j},$$
\item[]$\sigma_N^2=\frac{1}{w^2}\overline{\left\langle B^2\right\rangle}=\frac{1}{Pw^2}\sum_{\mu=1}^{P}\left\langle B^2\vert\mu\right\rangle$.
\end{itemize}
The relevant feature for our purposes is its limit behavior, as $N$ grows to infinity, with respect to $\alpha$: for $\alpha\geq\alpha_c$, $\lim_{N\to\infty}\frac{\sigma_N^2(y)}{N}\leq 1$, $\forall y$, while for $\alpha<\alpha_c$, $\lim_{N\to\infty}\frac{\sigma_N^2}{N^2}\leq k$, with $k$ constant (see figure \ref{surface3}, left). For more details see \cite{marsililibro}.

\subsection{Grand Canonical Minority Game}
The second market model we are going to analyze in this work is the GCMG (for more details see \cite{marsili_GCMG}). In this game each agent has only one trading strategy $a_i^{\mu(t)}=\pm w$ randomly chosen from the set of $2^P$ possible strategies. Each agent may decide not to play if the strategy is not good enough. More precisely, following the same formalism used to introduce the MG, each agent takes a bid $b_i(t)=\phi_i(t)a_i^{\mu(t)}$ where $\phi_i(t)=1$ or $0$ according to whether agent $i$ trades or not.

On the basis of the outcome $B(t)=\sum_{i=1}^Nb_i(t)=\sum_{i=1}^N\phi_i(t)a_{i}^{\mu(t)}$ each agent update his scores according to
\be
\label{increment_GCMG}
U_{i}(t+\delta t)=U_{i}(t)-a_{i}^{\mu(t)}\frac{B(t)}{P}-\frac{\epsilon_i}{P},
\ee
where $\delta t \to 0$ is the time increment.

If $-a_{i}^{\mu(t)}B(t)$ is larger than $\epsilon_i$, the score $U_i$ increases. The larger $U_i$ the more likely it is that the agent trades $(\phi_i(t)=1)$. Here we assume
$$
Prob\{\phi_i(t)=1\}=\frac{1}{1+e^{\Gamma U_{i}(t)}},
$$
with $\Gamma>0$.

The threshold $\epsilon_i$ models the incentives of agents for trading in the market; investors interested in trading because they need the market for exchanging assets have $\epsilon_i<0$ whilst speculators trading for profiting of price fluctuations have $\epsilon_i>0$. We will focus, for simplicity, on the case

\begin{eqnarray}
\epsilon_i&=&\epsilon>0 \quad {\rm for} \quad i\leq N_s\nonumber\\
\epsilon_i&=&-\infty \quad {\rm for} \quad N_s<i\leq N\nonumber.
\end{eqnarray}
The $N_p=N-N_s$ agents with $\epsilon_i=-\infty$ are the ${\it producers}$ trading no matter what, whereas $N_s$ are the speculators trades only if the cumulated performance of their active strategies increases more rapidly than $\epsilon t,$

From equation (4) and (5) in \cite{marsili_GCMG} the dynamics of the continuum time limit of the GCMG is given by the following $N$-dimensional stochastic differential equation
\begin{equation}
\label{stoch_eq1.1_GCMG}
dy_i(t)=\left(-\sum_{j=1}^N\overline{a_i b_j}H (y_j(t))-\epsilon\right)dt + A_i(\alpha,N,\Gamma,a,b)d{\mathbf W}(t), \quad i=1,\ldots,N
\end{equation}
where
\begin{itemize}
\item[]$y_i(t)=\Gamma U_i(\tau)$, $\tau=\frac{t}{\Gamma}$,
\item[]${\mathbf W}(t)$ is an $N$-dimensional Wiener process,
\item[]$A_i$ is the $i$-th row of the $N\times N$ matrix $A=(A_{ij})$ such that $$(AA')_{ij}(\alpha,N,\Gamma,a,b)=\frac{\Gamma\sigma_{N}^2}{P}\overline{a_i a_j}=\frac{\Gamma\sigma_{N}^2w^2(n_s+n_p)}{N}\overline{a_i a_j},$$
\item[]$\sigma_N^2=\frac{1}{w^2}\overline{\left\langle B^2\right\rangle}=\frac{1}{Pw^2}\sum_{\mu=1}^{P}\left\langle B^2\vert\mu\right\rangle$,
\item[]$H$ is the generalized Heaviside function ($Prob\{\phi_i(t)=0,1\}=H(y_i(t))$),
\item[]$\overline{(\ldots)}$ stands for average over $\mu(t)$.
\end{itemize}

\begin{figure}[h]
\centering
\includegraphics[width=0.5\textwidth,height=0.3\textwidth]{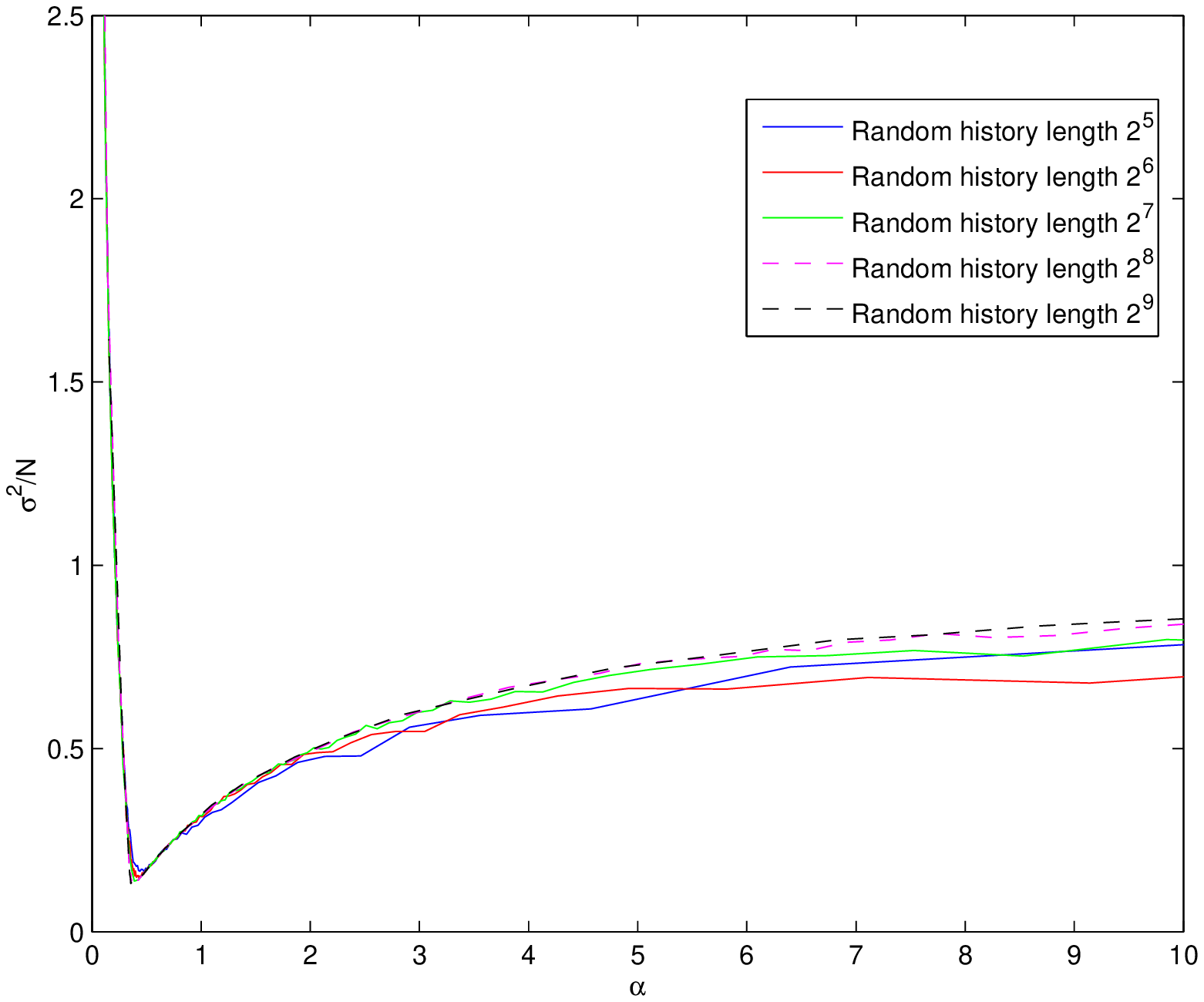}\includegraphics[width=0.5\textwidth,height=0.3\textwidth]{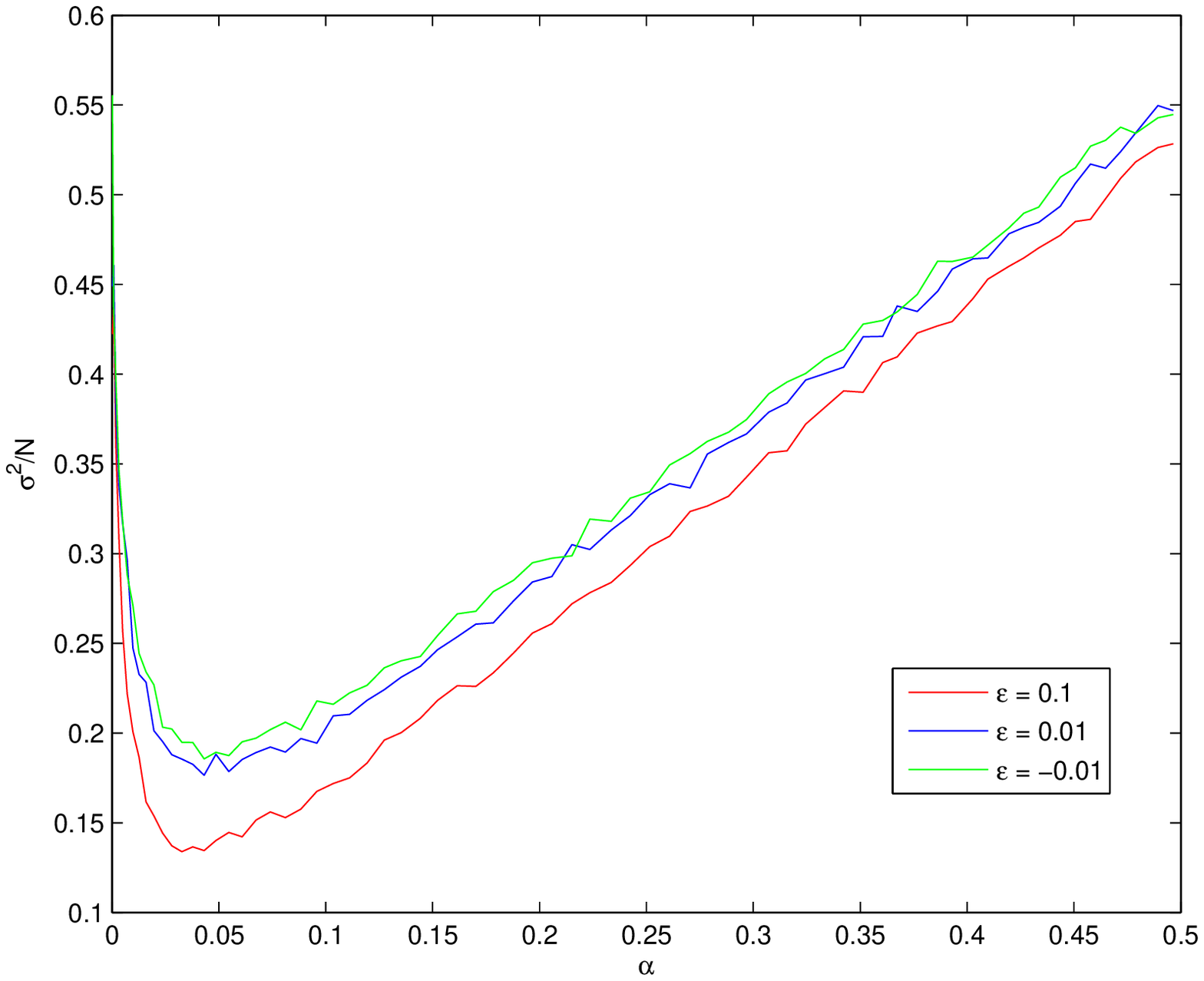}
\caption{Left: $\frac{\sigma^{2}_{N}}{N}$ as a function of $\alpha$ for the MG obtained numerically for different random histories. Right: $\frac{\sigma^{2}_{N}}{N}$ as a function of $\alpha_{n_s}$ for the GCMG obtained numerically for $L:=PN_s=8000$, $n_p=1$ and different values of $\epsilon$.}
\label{surface3}
\end{figure}

The relevant features of $\sigma_N^2$ for our purposes are in the thermodynamic limit, which is defined as the limit $N_s,N_p,P\to \infty$, keeping constant the fraction of speculators and producers $n_s=\frac{N_s}{P}$ and $n_p=\frac{N_p}{P}$. In this case the rule of key parameter $\alpha$ in the MG is played by $\alpha_{n_s}:=\frac{1}{n_s+n_p}$. The transition phase is marked by a critical value of speculators $\alpha_{n_s}^c(P)$: for $\alpha_{n_s}\geq \alpha_{n_s}^c(P)$, $\lim_{N\to	 \infty}\frac{\sigma_n^2}{N}(y)\leq 1 ~\forall y$, for $\alpha_{n_s}< \alpha_{n_s}^c(P)$, $\lim_{N\to	 \infty}\frac{\sigma_n^2}{N^2}(y)\leq k ~\forall y$ (see figure \ref{surface3}, right).

A key aspect that characterize the GCMG is that as soon as $\epsilon>0$ it is no more possible to find a solution for  $H_{\epsilon}=0$, where $H_{\epsilon}$ is the predictability function. Like it happens in the financial markets where $\epsilon$, sum of interest rate and transaction costs, represents incompressible costs, in the GCMG the onset of unpredictability corresponds to minima of $H_{\epsilon}$: as the fraction of speculators $n_s$ increases (and hence the system size increases) the minimum of volatility $\sigma_N^2$ decreases and the market becomes more and more unpredictable (see \cite{marsili_GCMG}).

\section{Stock price dynamics}
\label{stock}
In this section, using the scores difference stochastic differential equations, the overall market bid and hence the stock price dynamics is derived.

Let us consider a single stock in a market modeled by the continuum time MG or the continuum time GCMG;
following \cite{challet1}, we define the stock price dynamics in terms of excess demand, as
\be
\label{rel_1}
\log p(t+\delta t)=\log p(t) + \frac{B(t)}{Nw}
\ee
($Nw$ is the volume of trades and $w$ can be seen as the inverse liquidity parameter).

In the GCMG we are allowed to divide $B(t)$ not by the number of active traders $N_{\it act}$, but by the total number of agents $N$ because we are interested in results as $N\to \infty$ in the thermodynamic limit, where $N_p\to\infty$ and $N_s\to\infty$; hence as $N$ grows to infinity also the number of active traders grows to infinity of the same degree.

We address the analysis of the stock price dynamics for the MG and the GCMG separately.

\subsection{MG case}

Our aim is to obtain the continuum time dynamics of the stock price $p(t)$ from the continuum time dynamics of ${\mathbf y}(t)$ defined by (\ref{diff}). The key point is the following identity, that can be obtained directly from (\ref{increment})
$$
\frac{U_{+,i}(t+\delta t)-U_{-,i}(t+\delta t)}{2}=\frac{U_{+,i}(t)-U_{-,i}(t)}{2}-\frac{a_{+,i}^{\mu(t)}-a_{-,i}^{\mu(t)}}{2}\frac{B(t)}{P}.
$$
Taking into account relation (\ref{diff}) and that $\Gamma$ is finite, we obtain
$$
\frac{y_i(\Gamma t)}{\Gamma}-\frac{y_i(\Gamma (t+\delta t))}{\Gamma}=-\xi_i^{\mu(t)}\frac{B(t)}{\alpha N}.
$$
In the continuum time limit $\delta t\to 0$, 
\be
\label{eq_master}
dy_i(\Gamma t)=-\Gamma\xi_i^{\mu(t)}\frac{B(t)}{\alpha N}, \quad \forall i=1,2,\ldots,N.
\ee
Dynamics (\ref{eq_master}) reflects the stock price dynamics (\ref{rel_1}), where $\frac{B(t)}{N}$ is the generator of $\log p(t)$; indeed, like the right hand side of (\ref{rel_1}), also the right hand side of (\ref{eq_master}) does not contain a $dt$ term and $-\Gamma\xi_i^{\mu(t)}\frac{B(t)}{\alpha N}$ is the generator of $y_i(\Gamma t)$.

Multiplying both sides of (\ref{eq_master}) by $\xi_i^{\mu(t)}$, averaging over all the agents $i$, and taking the limit as $N$ goes to infinity we have
$$
\lim_{N\to\infty}\frac{1}{N}\sum_{i=1}^N\xi_i^{\mu(t)}dy_i(\Gamma t)=-\frac{\Gamma}{\alpha}\lim_{N\to\infty}\frac{1}{N}\sum_{i=1}^N\left(\xi_i^{\mu(t)}\right)^2\frac{B(t)}{N}.
$$ 
Since the $\xi_i^{\mu(t)}$ are independent from $B(t)$ (the difference between two actions given the past history $\mu(t)$ is fixed at the beginning of the game and does not depend on the action chosen by the agent on the basis of the score function $U_{s,i}(t)$), 
$$
\lim_{N\to\infty}\frac{1}{N}\sum_{i=1}^N\left(\xi_i^{\mu(t)}\right)^2\frac{B(t)}{N}
=\mathbb{E}\left[\left(\xi_i^{\mu(t)}\right)^2\right]\lim_{N\to\infty}\frac{B(t)}{N}=\frac{w^2}{2}wd\log p(t);
$$
it follows that
\be
\label{eq_price}
d \log p(t)=-\frac{2\alpha}{w^3\Gamma}\lim_{N\to\infty}\frac{1}{N}\sum_{i=1}^N \xi_i^{\mu(t)} dy_i (\Gamma t).
\ee
Integrating over $[0, t]$ both sides of (\ref{eq_price}) and taking into account equation (\ref{stoch_eq1.1}), it follows that
\begin{eqnarray}
\label{undiscounted_dynamics_old}
p(t)&=&p(0)\exp\left\{\lim_{N\to\infty} \int_0^t \frac{2\alpha}{w^3N}\sum_{i=1}^N\left[\xi_i^{\mu(s)}\left(\overline{\xi_i\Theta}+\sum_{j=1}^N\overline{\xi_i\xi_j}\tanh (y_j(\Gamma s))\right)\right]ds\right.\nonumber\\ 
&-&\left.\lim_{N\to\infty}\int_0^t \frac{2\alpha}{w^3N\Gamma}\sum_{i=1}^N\left(\xi_i^{\mu(s)}A_i\right)d{\mathbf W}(\Gamma s) \right\}.
\end{eqnarray}
The following proposition on the drift and diffusion term of equation (\ref{undiscounted_dynamics_old}) holds:

\begin{prop}
\label{prop_imp}
Let $t=O(N)$, that is $\lim_{N\to+\infty}\frac{t}{N}=k$. The drift term of (\ref{undiscounted_dynamics_old}) is at most $O(N)$; the diffusion term is different from zero $\forall \alpha$, finite for $\alpha\geq \alpha_c$ and at most $O(N)$ for $\alpha<\alpha_c$.
\end{prop}
\begin{flushleft}
\underline{Proof}
\end{flushleft}

It is easy to see that, by applying the Law of Large Numbers, $\forall i$ $\lim_{N\to\infty}\overline{\xi_i^2}=\frac{w^2}{2}$, $\lim_{N\to\infty}\overline{\xi_i\xi_j}=0$ and $\lim_{N\to\infty}\overline{\xi_i\Theta}=0$ (for detailed computations see \cite {ortisi}); it follows that
$$
\lim_{N\to\infty}\frac{2\alpha}{w^3N}\sum_{i=1}^N\left[\xi_i^{\mu(s)}\left(\overline{\xi_i\Theta}+\sum_{j=1}^N\overline{\xi_i\xi_j}\tanh (y_j(\Gamma s))\right)\right]=\frac{\alpha}{w^3}\lim_{N\to\infty}\frac{1}{N}\sum_{i=1}^N\xi_i^{\mu(s)}\tanh(y_i(\Gamma s))<k\alpha,
$$
with $k$ finite constant.
Hence the drift term is at most $k\alpha N$ and, due to different initial conditions and the stochastic evolution of $y_i$, in general different from zero.

Consider the zero mean random variable
$$
Y=\int_0^t \frac{2\alpha}{w^3N\Gamma}\sum_{i=1}^N\left(\xi_i^{\mu(s)}A_i\right)d{\mathbf W}(\Gamma s);
$$
since
\begin{equation}
\frac{2\alpha}{w^3N\Gamma}\sum_{i=1}^N\left(\xi_i^{\mu(s)}A_i\right)d{\mathbf W}(\Gamma s)=\sum_{j=1}^N\left(\frac{2\alpha}{w^3N\Gamma}\sum_{i=1}^N \xi_i^{\mu(s)}A_{ij} \right)dW_j(\Gamma s),\nonumber
\end{equation}
$Y$ has variance
\begin{eqnarray}
\label{eq_fund}
\nu&=&\int_0^t\sum_{j=1}^N\left(\frac{2\alpha}{w^3N\Gamma}\sum_{i=1}^N \xi_i^{\mu(s)}A_{ij} \right)^2d(\Gamma s)\nonumber\\
&=&\int_0^t\frac{4\alpha^2}{w^6N^2\Gamma}\sum_{i=1}^N\left(\xi_i^{\mu(s)}\right)^2(AA')_{ii}ds+\int_0^t\frac{4\alpha^2}{w^6N^2\Gamma}\sum_{i,j=1, i\neq j}^N\xi_i^{\mu(s)}\xi_j^{\mu(s)}(AA')_{ij}ds.
\end{eqnarray}
As $N$ grows to infinity, by applying the Law of Large Numbers, $\lim_{N\to\infty}AA'=\lim_{N\to\infty}\frac{\Gamma\sigma_N^2w^2}{2\alpha N}I$, where $I$ is the identity $N\times N$ matrix (see \cite{ortisi} for details), and the first term of the right hand side of (\ref{eq_fund}) is equal to $\lim_{N\to\infty}\frac{\alpha \sigma_N^2 t}{w^2N^2}$, while the second one is equal to $0$. Since for $\alpha\geq\alpha_c$, $0<\frac{\sigma_N^2}{N}<1$ and for $\alpha<\alpha_c$, $0<\frac{\sigma_N^2}{N}<kN$, 
the thesis follows.

\hspace{14cm}$\Box$

It is of worth to note that the hypothesis $t=O(N)$ is necessary to have the diffusion term different from zero and that it is the proportionality factor of the time needed to reach the stationary state of the MG. In numerical terms it means that, given a maturity $t$, a higher number of player needs a higher number of time steps, i.e. the player must interact more times to reach an equilibrium.

Consider the dynamics
\begin{eqnarray*}
q(t)&=&p(0)\exp\left\{ \int_0^{t} \frac{2\alpha}{w^3N}\sum_{i=1}^N\left[\xi_i^{\mu(s)}\left(\overline{\xi_i\Theta}+\sum_{j=1}^N\overline{\xi_i\xi_j}\tanh (y_j(\Gamma s))\right)\right]ds\right.\nonumber\\ 
&-&\left.\int_0^{t} \frac{2\alpha}{w^3N\Gamma}\sum_{i=1}^N\left(\xi_i^{\mu(s)}A_i\right)d{\mathbf W}(\Gamma s) \right\};
\end{eqnarray*}
by definition $q(t)$ converges pointwise towards $p(t)$ ($\lim_{N\to\infty}q(t)=p(t)$, $\forall \omega \in\Omega$) and hence almost surely; it follows that a.s. $\forall \epsilon_0 >0$ there exists $\overline{N}>0$ such that, $\forall N>\overline{N}$, $\vert p(t) - q(t)\vert<\epsilon_0$. By assuming $t=kN$ and $N$ finite but sufficiently large, we have that a.s. $q(t)\sim p(t)$ and proposition (\ref{prop_imp}) holds for $q(t)$.

From now on we assume $t=kN$ and $N$ finite but sufficiently large to have 
\begin{eqnarray}
\label{undiscounted_dynamics}
p(t)&=&p(0)\exp\left\{ \int_0^t \frac{2\alpha}{w^3N}\sum_{i=1}^N\left[\xi_i^{\mu(s)}\left(\overline{\xi_i\Theta}+\sum_{j=1}^N\overline{\xi_i\xi_j}\tanh (y_j(\Gamma s))\right)\right]ds\right.\nonumber\\ 
&&-\left.\int_0^t \frac{2\alpha}{w^3N\Gamma}\sum_{i=1}^N\left(\xi_i^{\mu(s)}A_i\right)d{\mathbf W}(\Gamma s) \right\}.
\end{eqnarray}
 
Let us define $\mathcal{G}_t=\sigma(W(\Gamma s), \xi^{\mu(s)}; s\leq t)$ the filtration generated by both the processes ${\mathbf W}$ and $\xi$. 
Since the drift and the diffusion terms are just the exponential of the drift and the diffusion of equation (\ref{stoch_eq1.1}) multiplied by $\xi_i^{\mu(s)}$, they are adapted to the filtration $\mathcal{G}_t$.

The differential of the stock price process is therefore, by Ito's formula,
\begin{eqnarray}
\frac{dp(t)}{p(t)}&=&\left[\frac{2\alpha}{w^3N}\sum_{i=1}^N\xi_i^{\mu(t)}\left(\overline{\xi_i\Theta}+\sum_{j=1}^N\overline{\xi_i\xi_j}\tanh (y_j(\Gamma t))\right)\right.\nonumber\\
&&\left.+\frac{2\alpha^2}{\Gamma w^6N^2}\left(\sum_{i=1}^N\xi_i^{\mu(t)}A_i\right)\left(\sum_{i=1}^N\xi_i^{\mu(t)}A_i\right)'\right]dt-\frac{2\alpha}{w^3N\Gamma}\sum_{i=1}^N\left(\xi_i^{\mu(t)}A_i\right)d{\mathbf W}(\Gamma t)\nonumber\\
&=&\left[\frac{2\alpha}{w^3N}\sum_{i=1}^N\xi_i^{\mu(t)}\left(\overline{\xi_i\Theta}+\sum_{j=1}^N\overline{\xi_i\xi_j}\tanh (y_j(\Gamma t))\right)+\frac{2\alpha^2}{w^6\Gamma N^2}\sum_{j=1}^N\left(\sum_{i=1}^N\xi_i^{\mu(t)}A_{ij}\right)^2\right]dt\nonumber\\
&&-\frac{2\alpha}{w^3N\Gamma}\sum_{i=1}^N\left(\xi_i^{\mu(t)}A_i\right)d{\mathbf W}(\Gamma t).\nonumber
\end{eqnarray}

\subsection{GCMG case}
For the case on the GCMG, since $y_i(t)=\Gamma U_i(\tau)$, it is immediate to see that for $\delta t\to 0$

\be
\label{eq_master_GCMG}
dy_i(\Gamma t)=-\frac{\Gamma}{P}\left( a_i^{\mu(t)}B(t)-\epsilon_i\right)=-\frac{\Gamma(n_s+n_p)}{N}\left( a_i^{\mu(t)}B(t)-\epsilon_i\right), \quad \forall i=1,2,\ldots,N.
\ee
Multiplying both sides of (\ref{eq_master_GCMG}) by $a_i^{\mu(t)}$, averaging over all the agents $i$, and taking the limit as $N$ goes to infinity we have
$$
\lim_{N\to\infty}\frac{1}{N}\sum_{i=1}^N a_i^{\mu(t)}dy_i(\Gamma t)=-\Gamma(n_s+n_p)\lim_{N\to\infty}\frac{1}{N}\sum_{i=1}^N\left(a_i^{\mu(t)}\right)^2\frac{B(t)}{N}+\Gamma(n_s+n_p)\lim_{N\to\infty}\frac{1}{N}\sum_{i=1}^N a_i(t)\frac{\epsilon_i}{N}.
$$
Since $a_i^{\mu(t)}$ have zero mean
$$
\lim_{N\to\infty}\frac{1}{N}\sum_{i=1}^N a_i(t)\frac{\epsilon_i}{N}=\lim_{N_s\to\infty}\frac{1}{N_s}\sum_{i=1}^{N_s} a_i(t)\frac{\epsilon}{N}+\lim_{N_p\to\infty}\frac{1}{N_p}\sum_{i=N_s+1}^{N} a_i(t)\frac{(-\infty)}{N}=0
$$ 
and are independent from $B(t)$,
$$
\lim_{N\to\infty}\frac{1}{N}\sum_{i=1}^N\left(a_i^{\mu(t)}\right)^2\frac{B(t)}{N}
=\mathbb{E}\left[\left(a_i^{\mu(t)}\right)^2\right]\lim_{N\to\infty}\frac{B(t)}{N}=w^3d\log p(t);
$$
it follows that
\be
\label{eq_price_GCMG}
d \log p(t)=-\frac{1}{w^3\Gamma(n_s+n_p)}\lim_{N\to\infty}\frac{1}{N}\sum_{i=1}^N a_i^{\mu(t)} dy_i (\Gamma t).
\ee
Since $n_s+n_p=\frac{N}{P}=\frac{1}{\alpha}$, we can put $\frac{1}{n_s+n_p}=\alpha_{n_s}$ ($n_p$ is itself a function of $n_s$). $\alpha_{n_s}$ as the same role of $\alpha$ in the MG, but whilst in the MG $n_p=0$ and hence $\alpha$ depends only on speculators, in the GCMG $n_s$ is the fundamental quantity  driving the behavior of $\sigma_N^2$.

Integrating over $[0, t]$ both sides of (\ref{eq_price_GCMG}) and taking into account equation (\ref{stoch_eq1.1_GCMG}), it follows that
\begin{eqnarray}
\label{undiscounted_dynamics_old_GCMG}
p(t)&=&p(0)\exp\left\{\lim_{N\to\infty} \int_0^t \frac{\alpha_{n_s}}{w^3N}\sum_{i=1}^N a_i^{\mu(s)} \left(\sum_{j=1}^N\overline{a_i b_j}H (y_j(\Gamma s))\right)ds\right.\nonumber\\ 
&-&\left.\lim_{N\to\infty}\int_0^t \frac{\alpha_{n_s}}{w^3N\Gamma}\sum_{i=1}^N\left(a_i^{\mu(s)}A_i\right)d{\mathbf W}(\Gamma s) \right\}.
\end{eqnarray}
Proposition \ref{prop_imp} holds also for (\ref{undiscounted_dynamics_old_GCMG}); the drift term is essentially the same of (\ref{undiscounted_dynamics_old}) and since in the GCMG $\lim_{N\to\infty}AA'=\lim_{N\to\infty}\frac{\Gamma \sigma_N^2w^2}{\alpha_{n_s}N}{\it I}$ proceeding in the same way we did for proving proposition \ref{prop_imp} we conclude that the diffusion term  is different from $0$, finite for $n_s\leq n_s^c(P)$ and at most $O(N)$ for $n_s> n_s^c(P)$. 

Also in this case we have that assuming $t=kN$ and $N$ finite but sufficiently large 
\begin{eqnarray}
\label{undiscounted_dynamics_GCMG}
p(t)&=&p(0)\exp\left\{ \int_0^t \frac{\alpha_{n_s}}{w^3N}\sum_{i=1}^N a_i^{\mu(s)} \left(\sum_{j=1}^N\overline{a_ib_j}H (y_j(\Gamma s))\right)ds\right.\nonumber\\ 
&&-\left.\int_0^t \frac{\alpha_{n_s}}{w^3N\Gamma}\sum_{i=1}^N\left(a_i^{\mu(s)}A_i\right)d{\mathbf W}(\Gamma s) \right\}.
\end{eqnarray}
 
Let us define $\mathcal{G}_t=\sigma(W(\Gamma s), a^{\mu(s)}; s\leq t)$ the filtration generated by both the processes ${\mathbf W}$ and $a$. 
Since the drift and the diffusion terms are just the exponential of the drift and the diffusion of equation (\ref{stoch_eq1.1_GCMG}) multiplied by $a_i^{\mu(s)}$, they are adapted to the filtration $\mathcal{G}_t$.

The differential of the stock price process is therefore, by Ito's formula,
\begin{eqnarray}
\frac{dp(t)}{p(t)}&=&\left[\frac{\alpha_{n_s}}{w^3N}\sum_{i=1}^Na_i^{\mu(t)}\left(\sum_{j=1}^N\overline{a_i b_j}H (y_j(\Gamma t))\right)\right.\nonumber\\
&&\left.+\frac{\alpha_{n_s}^2}{\Gamma w^6N^2}\left(\sum_{i=1}^Na_i^{\mu(t)}A_i\right)\left(\sum_{i=1}^Na_i^{\mu(t)}A_i\right)'\right]dt-\frac{\alpha_{n_s}}{w^3N\Gamma}\sum_{i=1}^N\left(a_i^{\mu(t)}A_i\right)d{\mathbf W}(\Gamma t)\nonumber\\
&=&\left[\frac{\alpha_{n_s}}{w^3N}\sum_{i=1}^Na_i^{\mu(t)}\left(\sum_{j=1}^N\overline{a_i b_j}H (y_j(\Gamma t))\right)+\frac{\alpha_{n_s}^2}{w^6\Gamma N^2}\sum_{j=1}^N\left(\sum_{i=1}^Na_i^{\mu(t)}A_{ij}\right)^2\right]dt\nonumber\\
&&-\frac{\alpha_{n_s}}{w^3N\Gamma}\sum_{i=1}^N\left(a_i^{\mu(t)}A_i\right)d{\mathbf W}(\Gamma t).\nonumber
\end{eqnarray}

\section{Derivative security pricing}
\label{risk-neutral}
In this section we develop the risk-neutral pricing of a derivative security on a single stock whose price dynamics is given by (\ref{undiscounted_dynamics}) and (\ref{undiscounted_dynamics_GCMG}).
We follow the usual scheme: construction of the risk-neutral measure, relying on Girsanov Theorem, for the discounted process and, by using the Martingale Representation Theorem, of the replication portfolio that allows to hedge a short position in the derivative security (see \cite{shreve}, p. 209-220).
At the end of the section we apply Black\&Scholes formula to price a European call option. 
\subsection{Discounted stock dynamics under risk-neutral: MG case}

Consider the interest rate process $R(s)$ adapted to the filtration $\mathcal{G}_t$, $0\leq t\leq T$. The discount process
$D(t)=e^{-\int_0^tR(s)ds}$ has, by Ito's formula, differential $dD(t)=-R(t)D(t)dt$.

The discounted stock process is
\begin{eqnarray*}
D(t)p(t)&=&p(0)\exp\left\{ \int_0^t-R(s)+ \frac{2\alpha}{w^3N}\sum_{i=1}^N\left[\xi_i^{\mu(s)}\left(\overline{\xi_i\Theta}+\sum_{j=1}^N\overline{\xi_i\xi_j}\tanh (y_j(\Gamma s))\right)\right]ds\right.\nonumber\\ 
&-&\left.\int_0^t \frac{2\alpha}{w^3N\Gamma}\sum_{i=1}^N\left(\xi_i^{\mu(s)}A_i\right)d{\mathbf W}(\Gamma s) \right\},
\end{eqnarray*}
and its differential
\begin{eqnarray*}
\frac{d(D(t)p(t))}{D(t)p(t)}&=&\left[\frac{2\alpha}{w^3N}\sum_{i=1}^N\left[\xi_i^{\mu(s)}\left(\overline{\xi_i\Theta}+\sum_{j=1}^N\overline{\xi_i\xi_j}\tanh (y_j(\Gamma s))\right)\right]\right.\nonumber\\
&&\left.+\frac{2\alpha^2}{w^6\Gamma N^2}\sum_{j=1}^N\left(\sum_{i=1}^N\xi_i^{\mu(t)}A_{ij}\right)^2-R(t)\right]dt
-\frac{2\alpha}{w^3N\Gamma}\sum_{i=1}^N\left(\xi_i^{\mu(t)}A_i\right)d{\mathbf W}(\Gamma t).\nonumber
\end{eqnarray*}

\begin{prop}
\label{measure}
If $T=O(N)$ there exists a probability measure $\widetilde{\mathbb{P}}$ (risk neutral measure) under which the discounted stock price $D(t)p(t)$ is a $\widetilde{\mathbb{P}}$-martingale.
\end{prop}
\begin{flushleft}
\underline{Proof}
\end{flushleft}
Let us define the market price of risk equation to be
\begin{eqnarray}
\frac{2\alpha}{w^3N\Gamma}\sum_{i=1}^N\left(\xi_i^{\mu(s)}A_i\right)\gamma(\Gamma t)&=&
\left[\frac{2\alpha}{w^3N}\sum_{i=1}^N\left[\xi_i^{\mu(s)}\left(\overline{\xi_i\Theta}+\sum_{j=1}^N\overline{\xi_i\xi_j}\tanh (y_j(\Gamma s))\right)\right]\right.+\nonumber\\
&&\left.+\frac{2\alpha^2}{w^6\Gamma N^2}\sum_{j=1}^N\left(\sum_{i=1}^N\xi_i^{\mu(t)}A_{ij}\right)^2-R(t)\right],\nonumber
\end{eqnarray}
where $\gamma(\Gamma t)=(\gamma_1(\Gamma t),\ldots,\gamma_N(\Gamma t))$ is an unknown process; the differential of the discounted process becomes
\be
d(D(t)p(t))=D(t)p(t)\frac{2\alpha}{w^3N\Gamma}\sum_{i=1}^N\left(\xi_i^{\mu(s)}A_i\right)[\gamma(\Gamma t)dt-d{\mathbf W}(\Gamma t)].\nonumber
\ee
Let $\vert\vert \cdot\vert\vert$ denote the usual Euclidean norm; we show that for $T=O(N)$, $\int_0^T\vert\vert\gamma(\Gamma u)\vert\vert^2d(\Gamma u)<\infty$.
Consider $N$ finite and sufficiently large (we remind that proposition (\ref{prop_imp}) for $N$ finite and sufficiently large); since, by proposition (\ref{prop_imp}) 
$$
\int_0^T\left\vert\left\vert\frac{2\alpha}{w^3N\Gamma}\sum_{i=1}^N\left(\xi_i^{\mu(u)}A_i\right)\right\vert\right\vert^2d(\Gamma u)=\int_0^T\sum_{j=1}^N\left(\frac{2\alpha}{w^3N\Gamma}\sum_{i=1}^N \xi_i^{\mu(u)}A_{ij} \right)^2d(\Gamma u)>0 {\rm ~and~ finite}, 
$$
$$
\int_0^T\left\vert\left\vert\frac{2\alpha}{w^3N}\sum_{i=1}^N\left[\xi_i^{\mu(s)}\left(\overline{\xi_i\Theta}+\sum_{j=1}^N\overline{\xi_i\xi_j}\tanh (y_j(\Gamma s))\right)\right]\right\vert\right\vert^2d(\Gamma u)<\infty,
$$
and
$$
\int_0^T\left\vert\left\vert \frac{2\alpha^2}{w^6\Gamma N^2}\sum_{j=1}^N\left(\sum_{i=1}^N\xi_i^{\mu(t)}A_{ij}\right)^2\right\vert\right\vert^2d(\Gamma u)\sim 0,
$$
we have
$$
0<\int_0^T\left\vert\left\vert\gamma(\Gamma u)\right\vert\right\vert^2d(\Gamma u)<\infty.
$$
It follows that $\gamma(\Gamma t)$ satisfies the Novikov condition (see \cite{nov})
$$
\mathbb{E}\left[\frac{1}{2}\int_0^T\vert\vert\gamma(\Gamma u)\vert\vert^2du\right]<\infty.
$$
As a consequence we can apply Girsanov's Theorem (see \cite{kar}, p.198); there exists a probability measure $\widetilde{\mathbb{P}}$ (risk neutral measure) under which $\widetilde{{\mathbf W}}(\Gamma t)=\int_0^t\gamma(\Gamma u)du-{\mathbf W}(\Gamma t)$ is an $N$-dimensional Brownian motion.
It follows that
$$
d(D(t)p(t))=D(t)p(t)\frac{2\alpha}{w^3N\Gamma}\sum_{i=1}^N\left(\xi_i^{\mu(t)}A_i\right)d\widetilde{{\mathbf W}}(\Gamma t),
$$
and hence the discounted stock price is a $\widetilde{\mathbb{P}}$-martingale.

\hspace{14cm}$\Box$

By making the replacement $dW(\Gamma t)=\gamma(\Gamma t)dt-d\widetilde{W}(\Gamma t)$, dynamics (\ref{undiscounted_dynamics}) of the undiscounted stock price $p(t)$ becomes
\begin{eqnarray}
\label{stock_dynamics}
p(t)&=&p(0)\exp\left\{ \int_0^t R(s)-\frac{2\alpha^2}{w^6\Gamma N^2}\sum_{j=1}^N\left(\sum_{i=1}^N\xi_i^{\mu(t)}A_{ij}\right)^2ds+\right.\nonumber\\
&&\left.\int_0^t \frac{2\alpha}{w^3N\Gamma}\sum_{i=1}^N\left(\xi_i^{\mu(s)}A_i\right)d\widetilde{{\mathbf W}}(\Gamma s) \right\}.
\end{eqnarray}
\subsection{Discounted stock dynamics under risk-neutral: GCMG case}

Consider the interest rate process $R(s)$ adapted to the filtration $\mathcal{G}_t$, $0\leq t\leq T$. The discount process
$D(t)=e^{-\int_0^tR(s)ds}$ has, by Ito's formula, differential $dD(t)=-R(t)D(t)dt$.

The discounted stock process is
\begin{eqnarray*}
D(t)p(t)&=&p(0)\exp\left\{ \int_0^t-R(s)+ \frac{\alpha_{n_s}}{w^3N}\sum_{i=1}^Na_i^{\mu(s)}\left(\sum_{j=1}^N\overline{a_i b_j}H (y_j(\Gamma s))\right)ds\right.\nonumber\\ 
&-&\left.\int_0^t \frac{\alpha_{n_s}}{w^3N\Gamma}\sum_{i=1}^N\left(a_i^{\mu(s)}A_i\right)d{\mathbf W}(\Gamma s) \right\},
\end{eqnarray*}
and its differential
\begin{eqnarray*}
\frac{d(D(t)p(t))}{D(t)p(t)}&=&\left[\frac{\alpha_{n_s}}{w^3N}\sum_{i=1}^Na_i^{\mu(s)}\left(\sum_{j=1}^N\overline{a_i b_j}H (y_j(\Gamma s))\right)\right.\nonumber\\
&&\left.+\frac{\alpha_{n_s}^2}{w^6\Gamma N^2}\sum_{j=1}^N\left(\sum_{i=1}^N a_i^{\mu(t)}A_{ij}\right)^2-R(t)\right]dt
-\frac{\alpha_{n_s}}{w^3N\Gamma}\sum_{i=1}^N\left(a_i^{\mu(t)}A_i\right)d{\mathbf W}(\Gamma t).\nonumber
\end{eqnarray*}

With the natural modifications to the market price of risk equation in proposition \ref{measure}, the same proposition holds also for the GCMG and dynamics (\ref{undiscounted_dynamics_GCMG}) of the undiscounted stock price $p(t)$ becomes
\begin{eqnarray}
\label{stock_dynamics_GCMG}
p(t)&=&p(0)\exp\left\{ \int_0^t R(s)-\frac{\alpha_{n_s}^2}{w^6\Gamma N^2}\sum_{j=1}^N\left(\sum_{i=1}^N a_i^{\mu(s)}A_{ij}\right)^2ds+\right.\nonumber\\
&&\left.\int_0^t \frac{\alpha_{n_s}}{w^3N\Gamma}\sum_{i=1}^N\left(a_i^{\mu(s)}A_i\right)d\widetilde{{\mathbf W}}(\Gamma s) \right\}.
\end{eqnarray}

\subsection{European call option pricing}

Since, for $T-t=O(N)$, there exists a risk neutral measure $\widetilde{\mathbb{P}}$ for the discounted process $D(t)p(t)$, we can apply the usual scheme, relying on the martingale representation theorem, to show the existence of a replication portfolio for a derivative security pricing.

Consider $T-t=O(N)$;
let $V(T)$ be an $\mathcal{G}_T$-measurable random variable representing the payoff at time $T$ of a derivative security. The process $E(t)=\mathbb{E}_{\widetilde{\mathbb{P}}}[D(T)V(T)\vert \mathcal{G}_t]$ is a $\widetilde{\mathbb{P}}$-martingale (it follows from iterated conditioning); by the Martingale Representation Theorem there exists an initial capital $X(0)$ and a portfolio strategy $\Delta(t)$ such that $X(T)=V(T)$ almost surely and an adapted process $\phi(t)$ which constructs $E(t)$ out of $D(t)p(t)$. It follows that $D(t)X(t)$ is a $\widetilde{\mathbb{P}}$-martingale
\be
\label{hedging}
D(t)X(t)=\mathbb{E}_{\widetilde{\mathbb{P}}}[D(T)X(T)\vert \mathcal{G}_t]=\mathbb{E}_{\widetilde{\mathbb{P}}}[D(T)V(T)\vert \mathcal{G}_t].
\ee
The value $X(t)$ of the hedging portfolio in (\ref{hedging}) is the capital needed at time $t$ in order to successfully hedge the position in the derivative security with payoff $V(T)$. Hence $V(t)$ is the price of the derivative security at time $t$ and we obtain the usual risk neutral pricing formula (see \cite{shreve}, p.218-222)
$$
D(t)V(t)=\mathbb{E}_{\widetilde{\mathbb{P}}}[D(T)V(T)\vert \mathcal{G}_t], \quad 0\leq t\leq T;
$$
by recalling the definition of $D(t)$,
$$
\label{pricing_formula}
V(t)=\mathbb{E}_{\widetilde{\mathbb{P}}}[e^{-\int_t^TR(s)ds}V(T)\vert \mathcal{G}_t], \quad 0\leq t\leq T.
$$

\begin{prop}
Let us assume constant interest rate $r$; the price of a European call option with underlying $p(t)$ and strike $K$ is
\be
\label{pricing}
c(t,p(t);K,r,\nu)=p(t)\psi(d(\theta,p(t)))-e^{-r\theta}K\psi \left(d\left(\theta,p(t)\right)- \sqrt{\nu}\sqrt{\theta}\right),\nonumber
\ee
where $\theta=T-t$, $\nu=\nu_{\rm MG}=\frac{\alpha\sigma^2_N}{w^2N^2}$ for the MG, $\nu=\nu_{\rm GCMG}=\frac{\alpha_{n_s}\sigma^2_N}{w^2N^2}$ for the GCMG, $d(\theta,p(t))=\frac{\left[\log\frac{p(t)}{K}+\left( r + \frac{1}{2}\nu \right)\theta \right]}{\sqrt{\nu\theta}}$ and $\psi$ the ${\mathrm erf}$ function.
\end{prop}
\begin{flushleft}
\underline{Proof}
\end{flushleft}
Consider the random variable
$$
Y=\int_t^T \frac{2\alpha}{w^3N\Gamma}\sum_{i=1}^N\left(\xi_i^{\mu(s)}A_i\right)d\widetilde{{\mathbf W}}(\Gamma s)=\sum_{j=1}^N\int_t^T\frac{2\alpha}{w^3N\Gamma}\sum_{i=1}^N \left(\xi_i^{\mu(s)}A_{ij} \right)dW_j(\Gamma s)
$$
for the MG case and
$$
Y=\int_t^T \frac{\alpha_{n_s}}{w^3N\Gamma}\sum_{i=1}^N\left(a_i^{\mu(s)}A_i\right)d\widetilde{{\mathbf W}}(\Gamma s)=\sum_{j=1}^N\int_t^T\frac{\alpha_{n_s}}{w^3N\Gamma}\sum_{i=1}^N \left(a_i^{\mu(s)}A_{ij} \right)dW_j(\Gamma s)
$$
for the GCMG case.

$Y$ has zero mean and, in the limit as $N$ grows to infinity, finite variance $\nu$ (see prop. \ref{prop_imp}); since we consider $N$ sufficiently large, we can assume ${\mathrm Var[Y]}=\nu$. For $N$ sufficiently large we can apply the results obtained at the end of proof of proposition (\ref{prop_imp}); it follows that $Y$ has a normal distribution with $0$ mean and variance $\nu_{\rm MG}=\frac{\alpha\sigma_N^2\theta}{w^2N^2}$, $\nu_{\rm GCMG}=\frac{\alpha_{n_s}\sigma_N^2\theta}{w^2N^2}$ and equations (\ref{stock_dynamics}) and (\ref{stock_dynamics_GCMG}) become
\begin{equation}
p(T)=p(t)\exp\left\{ \left(r-\frac{1}{2}\frac{\nu}{\theta}\right)\theta-\sqrt{\frac{\nu}{\theta}}\sqrt{\theta}Z \right\},\nonumber
\end{equation}
$Z=-\frac{Y}{\sqrt{\nu}}$ is a standard normal random variable.

We can apply Black\&Scholes formula (see \cite{shreve}, pp.218-220) and obtain the price of a call option with underlying $p$ and strike $K$:
$$
c(t,p(t);K,r,\nu)=p(t)\psi(d(\theta,p(t)))-e^{-r\theta}K\psi \left(d\left(\theta,p(t)\right)- \sqrt{\nu}\right),
$$
where $d(\theta,p(t))=\frac{\left[\log\frac{p(t)}{K}+\left( r + \frac{1}{2}\frac{\nu}{\theta} \right)\theta \right]}{\sqrt{\nu}}$ and $\psi$ the {\it erf} function.

\hspace{14cm}$\Box$

\section{Model calibration}
\label{section_5}
Calibrating the model means to find values $\overline{w}$, $\overline{\alpha}$,$\overline{\alpha}_{n_s}$, $\overline{\sigma_N^2}$ such that the option implied volatility available on the market $\sqrt{\nu^{\it impl}}$ equals $\sqrt{\nu}$, where $\sqrt{\nu}$ is the implied volatility coming from the European call option pricing formula (\ref{pricing}). Whilst parameters $\alpha$, $\alpha_{n_s}$ and $\sigma_N^2$ are characteristic of the model, parameter $w$ is related to the market liquidity of the stock. As a consequence it is meaningful to divide the calibration in two steps. The first step consists in finding a value $\overline{w}$ to link the stock price return volatility with the market volatility $\sqrt{\nu^{\it mkt}}$; the second step consists in finding, given the value of $\overline{w}$ already calibrated, values $\overline{\alpha}$, $\overline{\alpha}_{n_s}$ and $\overline{\sigma_N^2}$ such that $\nu^{\it impl}=\nu$.

At this point it worths to define the market volatility $\sqrt{\nu^{\it mkt}}$. The market volatility is an index representing the volatility present in the market. It could be the historic volatility of the returns of an equity index, like the S\&P500 or the DAX; in this case we should choose the time window to compute the historic volatility (1 month, 1 year,...). The other option is to choose an index like the VIX or the VDAX; these indices, widely used by finance practitioners as a measure of the risk present in the market, are built starting from listed options on S\&P500 or DAX and have the advantage of being uniquely defined (no need to define a time window), more responsive to change in market regime than the historic volatility and of having almost the same behavior of the historic ones (maxima and minima of the short term historic volatility correspond with high degree to the maxima and minima of the implied one). It follows that for our calibration purpose we can use VDAX index as market volatility: what we need is just to fix values for $(\alpha,\sigma_N^2)$ and $(\alpha_{n_s},\sigma_N^2)$ respectively and find a value for $\omega$ such that $\sqrt{\nu}=\sqrt{\nu^{\it mkt}}$.
Since the stock price return variance is ${\mathrm Var}\left[\log\frac{p(t+\delta t)}{p(t)}\right]=\frac{\sigma_N^2}{N}\frac{1}{w^2N}$, it reaches the minimum value when $\sigma_N^2=\sigma_c^2$; it follows that also the minimum of the market volatility $\sqrt{\nu^{\it mkt}}$ corresponds to the critical value $\sigma_c$. This requires the market to be non-stationary. From the market point of view this assumption is natural: since stationarity corresponds to the transition point between efficiency and inefficiency it is fair to consider  the market efficient or inefficient, but not at the critical point; from the game point of view we have to require the game parameters not to be at the critical point $\alpha_c$.

Taking into account the expression of the implied volatility $\nu$, $\overline{w}$ is such that
\begin{equation}
\label{mkt_equation}
\sqrt{\bar{\nu}^{mkt}}=\sqrt{\alpha_c}\frac{\sigma_c}{\bar{w}N},
\end{equation}
where $\sqrt{\bar{\nu}^{mkt}}$ is the minimum market volatility.

The second step consists in calibrating the MG and the GCMG on the options, that is to find couples $(\overline{\alpha},\overline{\sigma_N^2})$ and $(\overline{\alpha}_{n_s},\overline{\sigma_N^2})$ respectively such that:
$$
\sqrt{\nu^{\it impl}}=\sqrt{\overline{\alpha}}\frac{\overline{\sigma_{N}}}{\bar{w}N}
$$
for the MG and
$$
\sqrt{\nu^{\it impl}}=\sqrt{\overline{\alpha}_{n_s}}\frac{\overline{\sigma_{N}}}{\bar{w}N}
$$
for the GCMG, where $\sqrt{\nu^{\it impl}}$ is the option implied volatility available on the market.

In order to perform the calibration on the real market we have first of all to select a stock. 
We choose the DAX Index, a total return index reinvesting dividends and representing the most relevant stocks traded at the German stocks exchange, hence a very liquid instrument with a long time series. The calibration of $\bar{w}$ is performed over VDAX Index, which is the index of implied volatility for DAX options. 
The minimum volatility, i.e. the minimum value of VDAX, since $03/01/2000$ is reached the $11/02/2005$ with a value of $10.98\%$: $\sqrt{\overline{\nu}^{\it mkt}}=10.98\%$. For both the MG and the GCMG $\overline{w}$ can be obtained from the market volatility in correspondence with the minumun value $\sigma_c$ reached by $\sigma_N$; of course the $\sigma_c$ of the MG is different from the one of the GCMG. The next calibration, performed for the MG as $M$ goes from 5 to 9 (it is not possible to calibrate for $M$ too small), and for the GCMG for different values of $\epsilon$ and $L:=PN_s$, is ran for all options once. 

\begin{figure}[h]
\centering
\includegraphics[width=0.5\textwidth,height=0.32\textwidth]{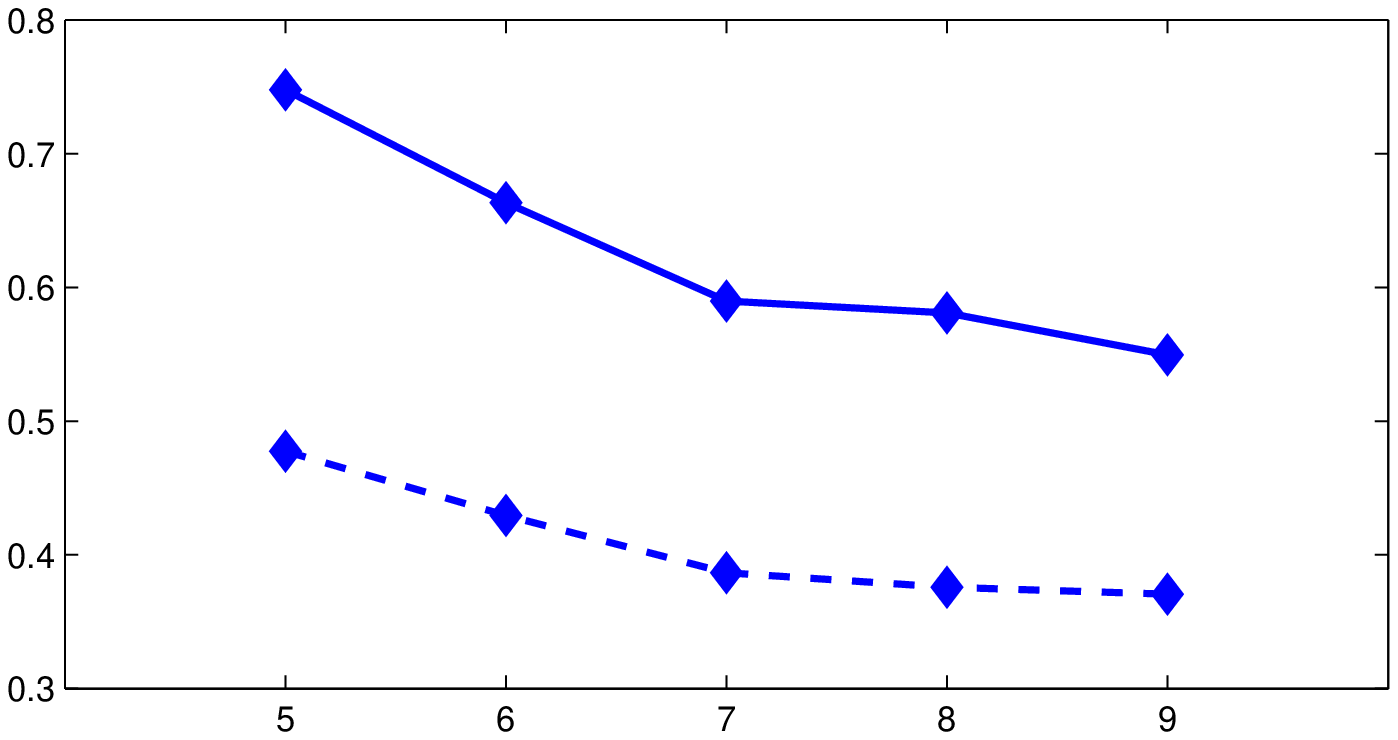}\includegraphics[width=0.5\textwidth]{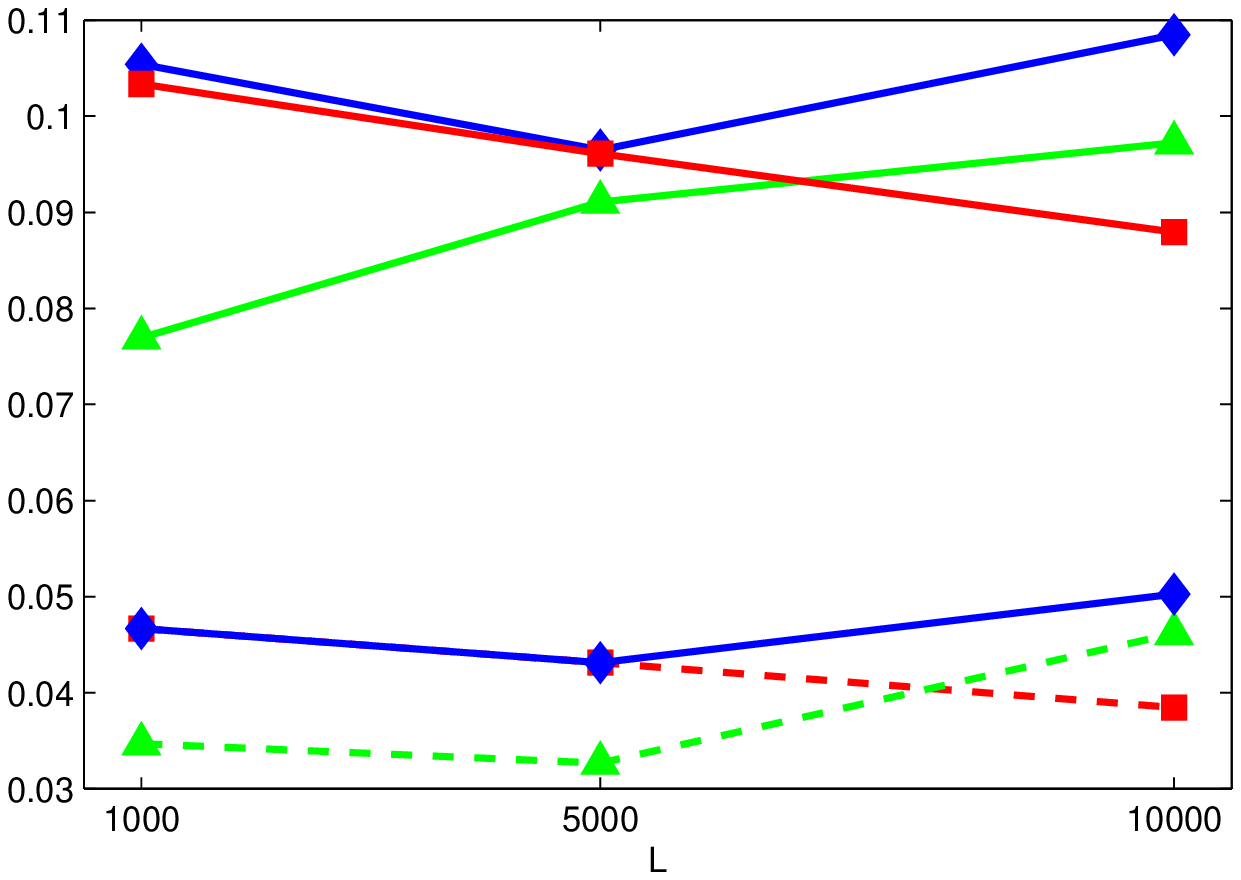}
\caption{Left: Comparison between average calibrated $\alpha$ (continuos line) and average $\alpha_c$ (dashed line) for different values of $M$. Right: Comparison between average calibrated $\alpha_{n_s}$ (continuos line) and average $\alpha_{n_s}^c$(dashed line) for different values of $L$ and $\epsilon=0.01$ (diamonds), $\epsilon=-0.01$ (squares), $\epsilon=0.1$ (triangles).}
\label{figure_alpha_ratios_MG}
\end{figure}

\begin{figure}[h]
\centering
\includegraphics[width=0.5\textwidth,height=0.3\textwidth]{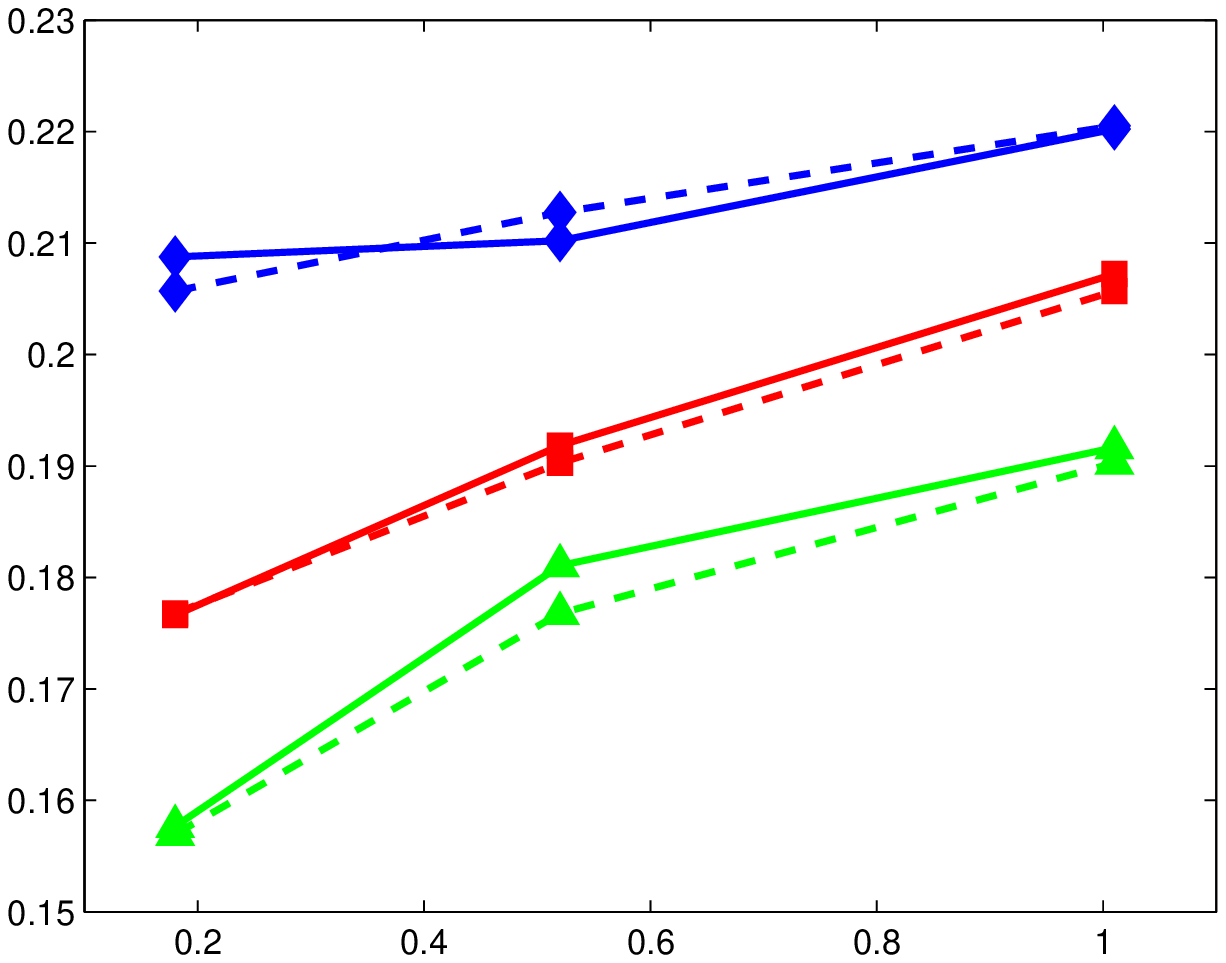}\includegraphics[width=0.5\textwidth]{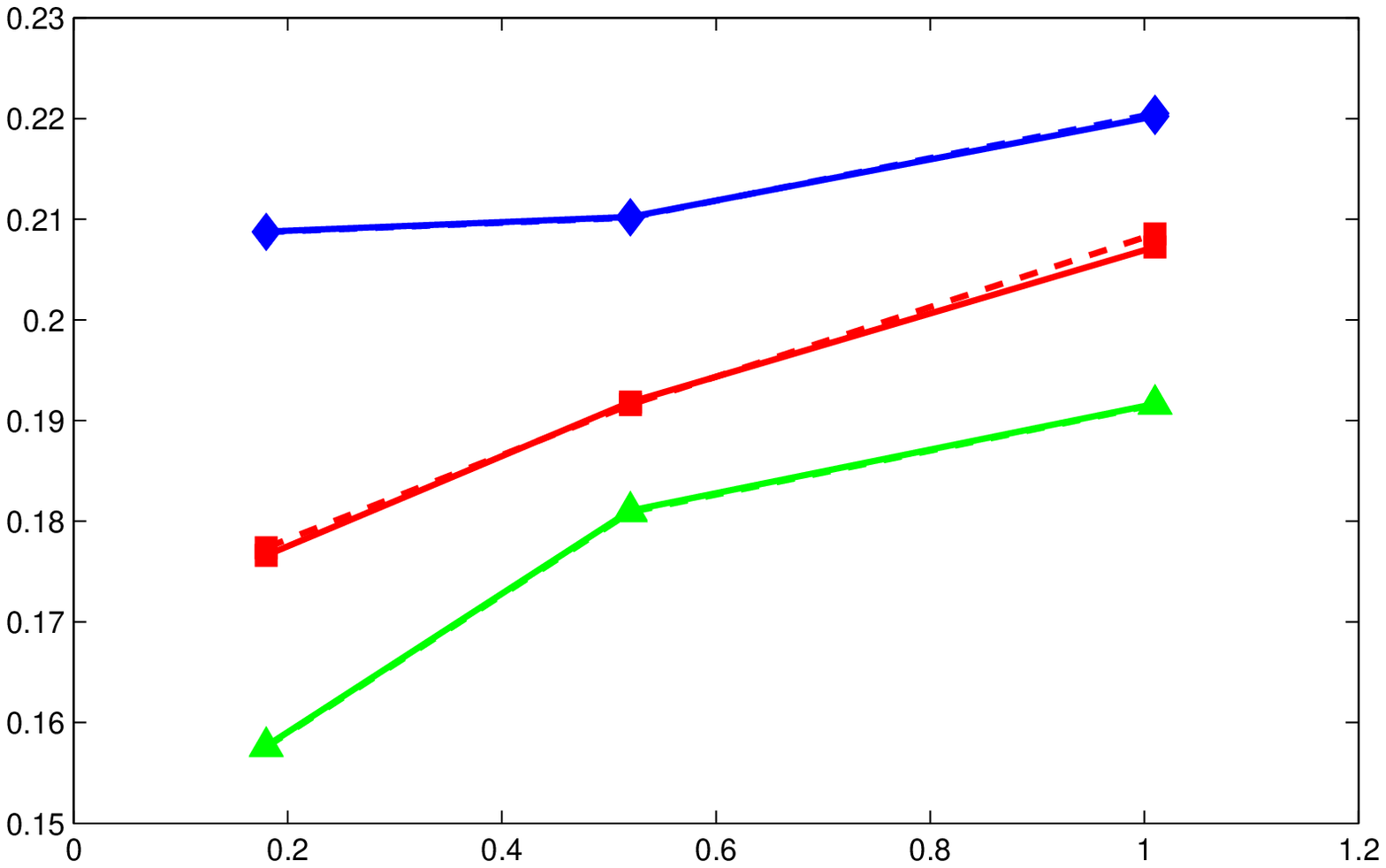}
\caption{Comparison between market option implied volatility (continuous line) and MG implied volatility (dashed line). Horizontal axis displays option time to maturity, vertical axis volatility level for options with moneyness equal to 1.06 (diamonds), 1 (squares), 0.94 (triangles). Left: $M=6$, right: $M=9$.}
\label{figure_options_MG_6}
\end{figure}

\begin{figure}[h]
\centering
\includegraphics[width=0.5\textwidth]{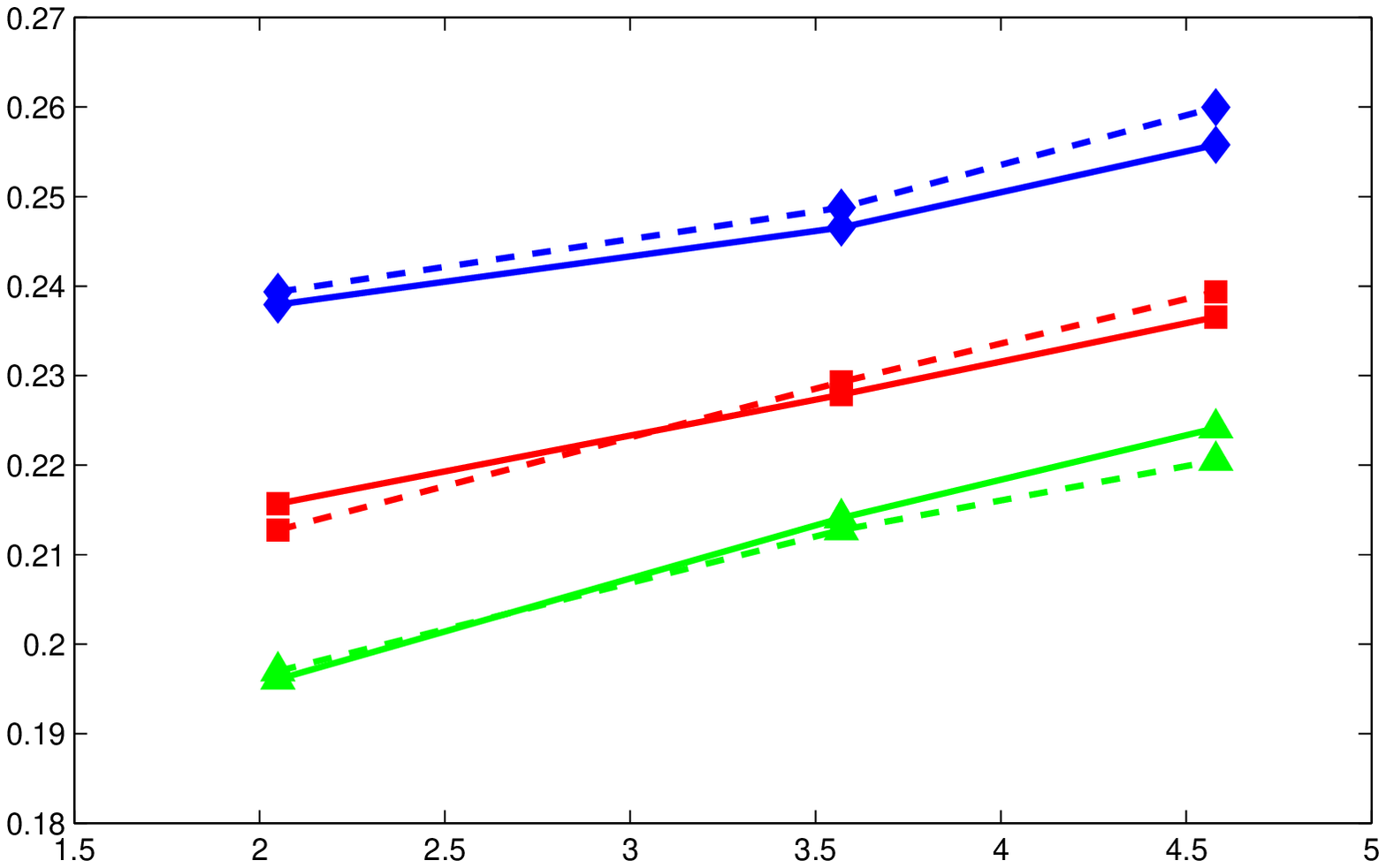}\includegraphics[width=0.5\textwidth]{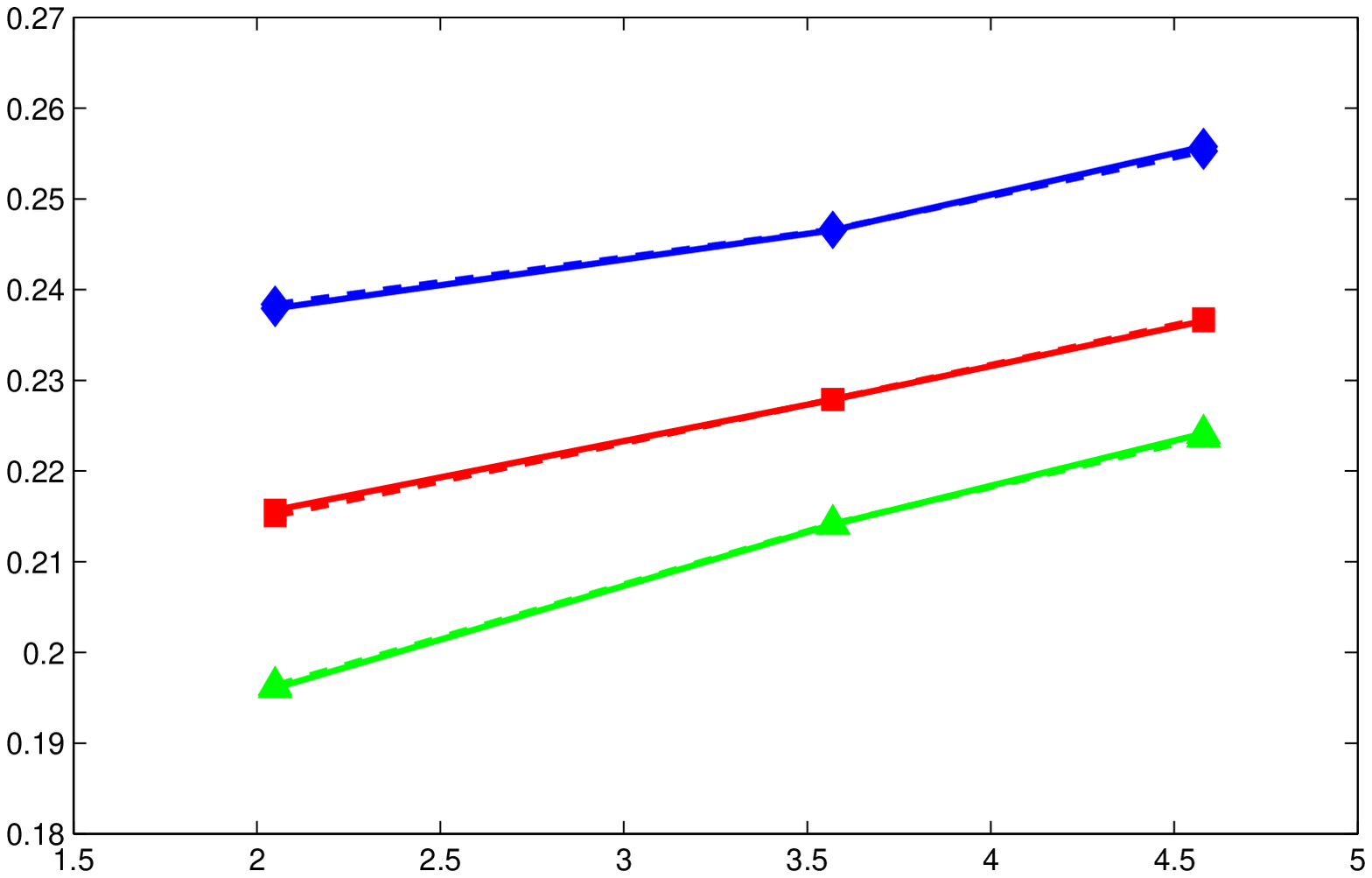}
\caption{Comparison between market option implied volatility (continuous line) and MG implied volatility (dashed line). Horizontal axis displays option time to maturity, vertical axis volatility level for options with moneyness equal to 1.1 (diamonds), 0.98 (squares), 0.88 (triangles). Left: $M=6$, right: $M=9$.}
\label{figure_options_MG_9}
\end{figure}

\begin{figure}[h]
\centering
\includegraphics[width=0.5\textwidth]{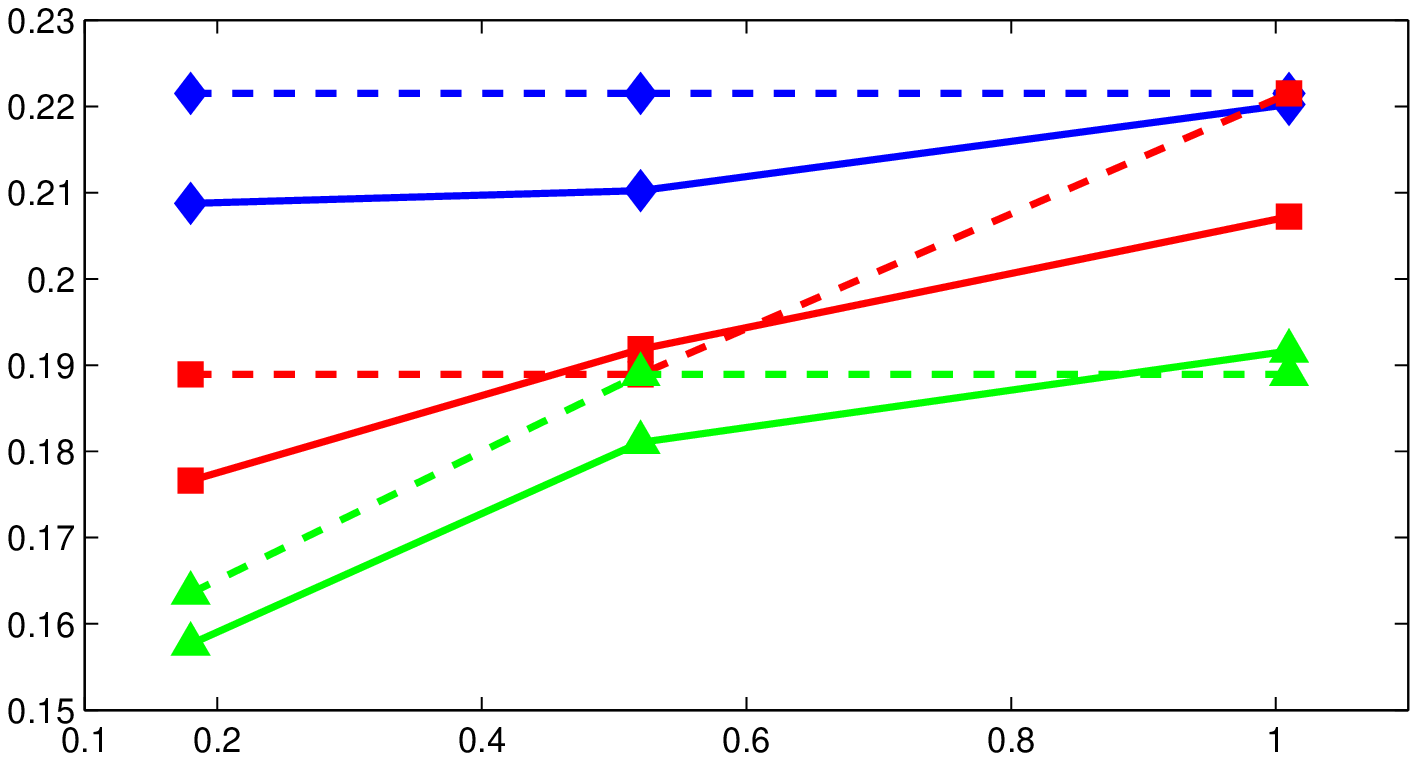}\includegraphics[width=0.5\textwidth]{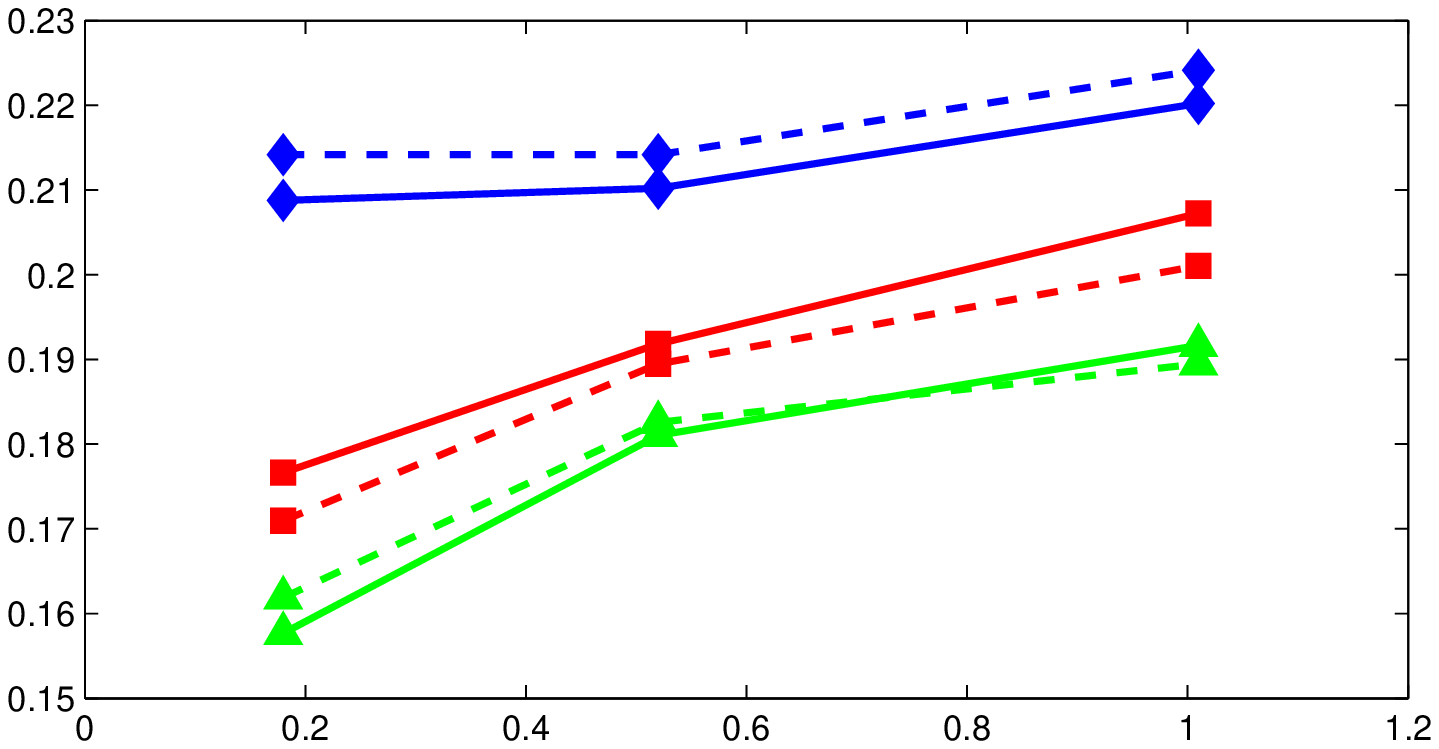}
\caption{Comparison between market option implied volatility (continuous line) and GCMG implied volatility (dashed line). Horizontal axis displays option time to maturity, vertical axis volatility level for options with moneyness equal to 1.06 (diamonds), 1 (squares), 0.94 (triangles). Left: $L=1000$, $\epsilon=0.01$, right: $L=10000$, $\epsilon=0.01$.}
\label{figure_options_GMG_1}
\end{figure}

\begin{figure}[h]
\centering
\includegraphics[width=0.5\textwidth,height=0.26\textwidth]{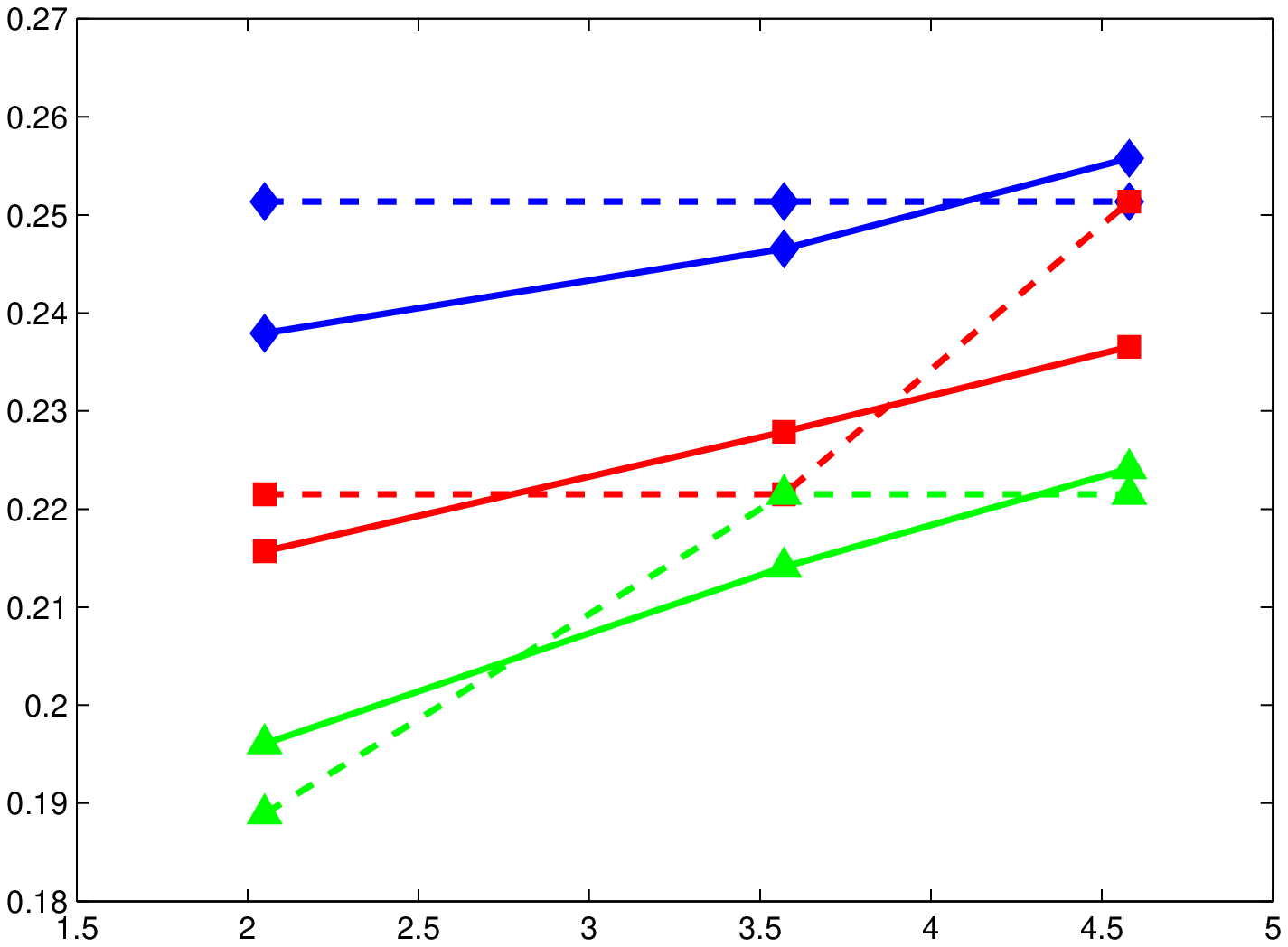}\includegraphics[width=0.5\textwidth]{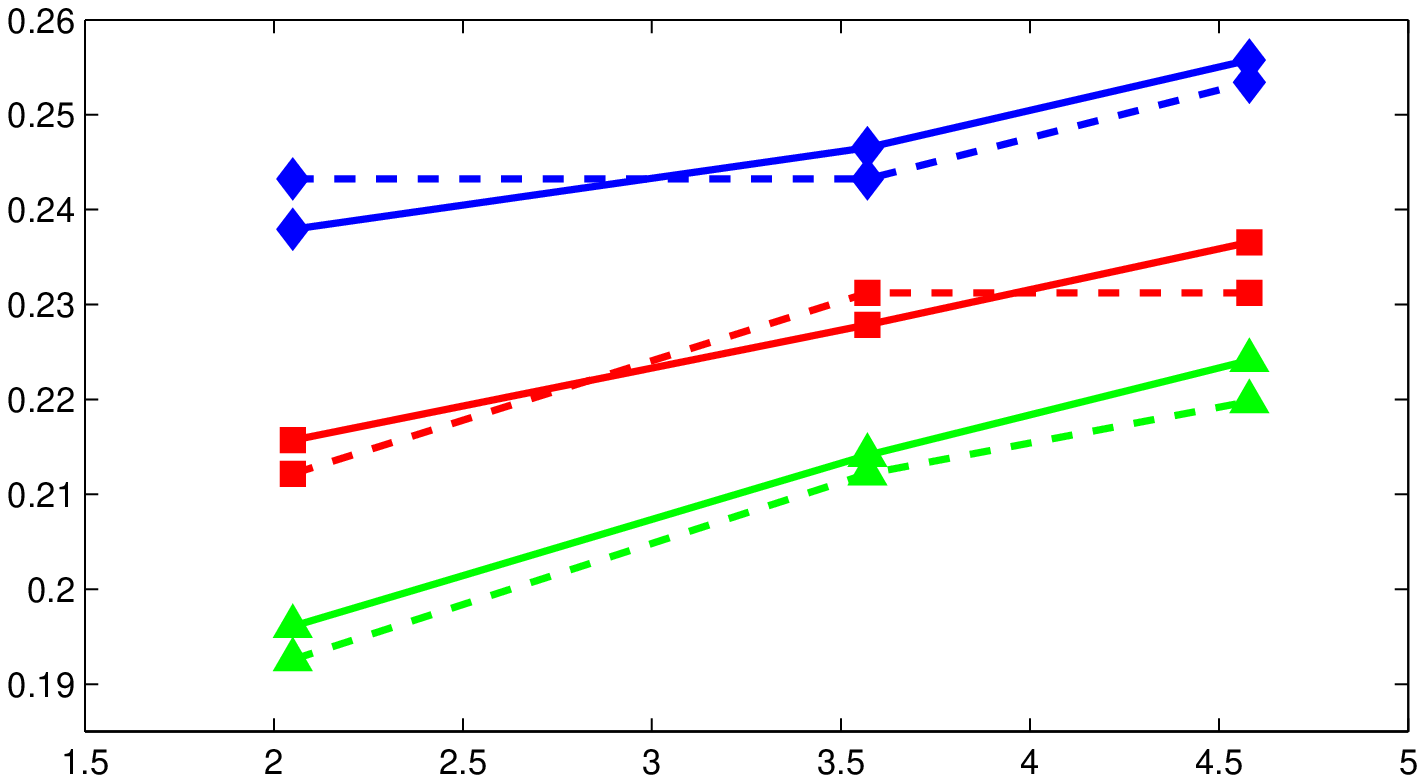}
\caption{Comparison between market option implied volatility (continuous line) and GCMG implied volatility (dashed line). Horizontal axis displays option time to maturity, vertical axis volatility level for options with moneyness equal to 1.1 (diamonds), 0.98 (squares), 0.88 (triangles). Left: $L=1000$, $\epsilon=0.01$, right: $L=10000$, $\epsilon=0.01$.}
\label{figure_options_GMG_2}
\end{figure}

\begin{figure}[h]
\centering
\includegraphics[width=0.5\textwidth]{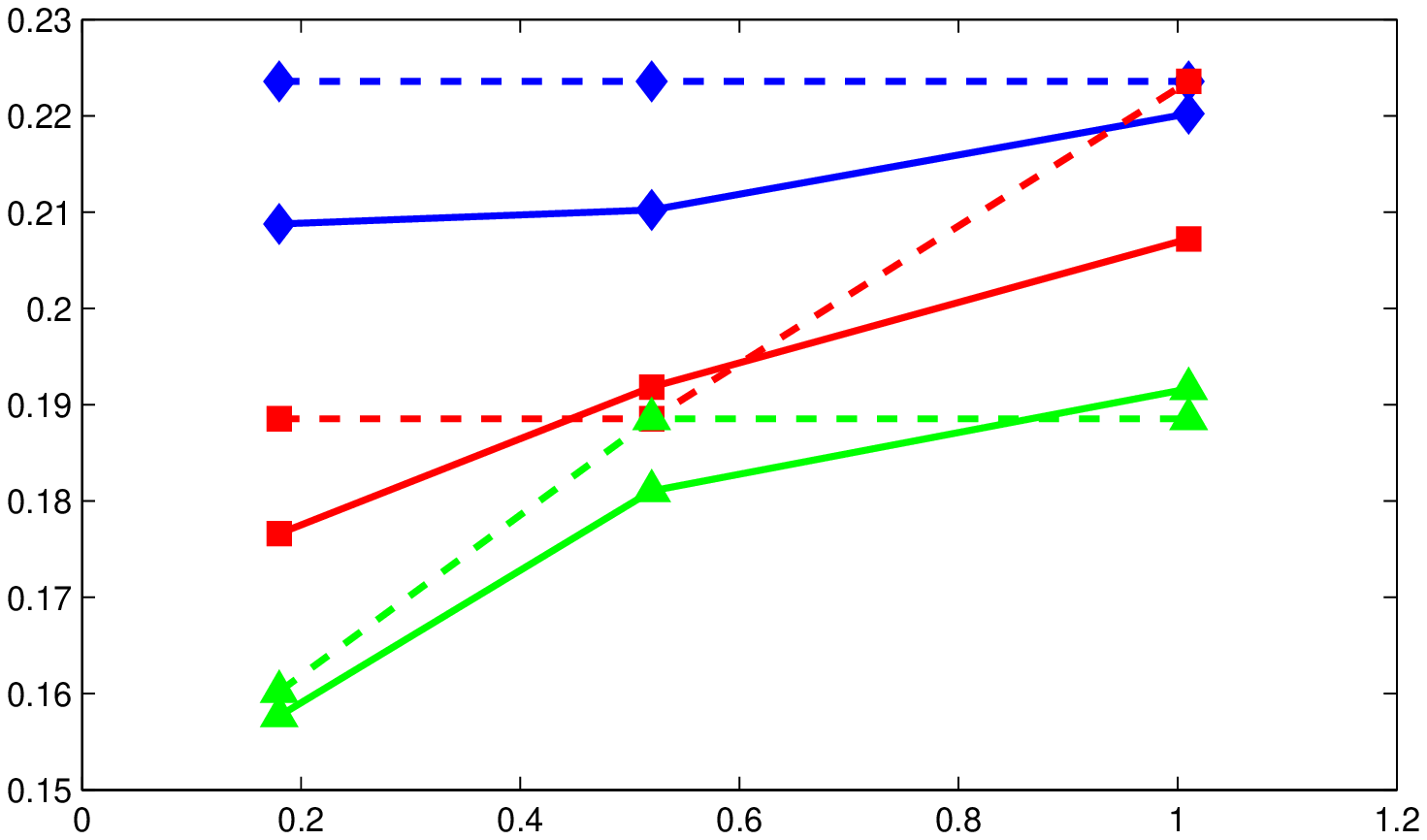}\includegraphics[width=0.5\textwidth]{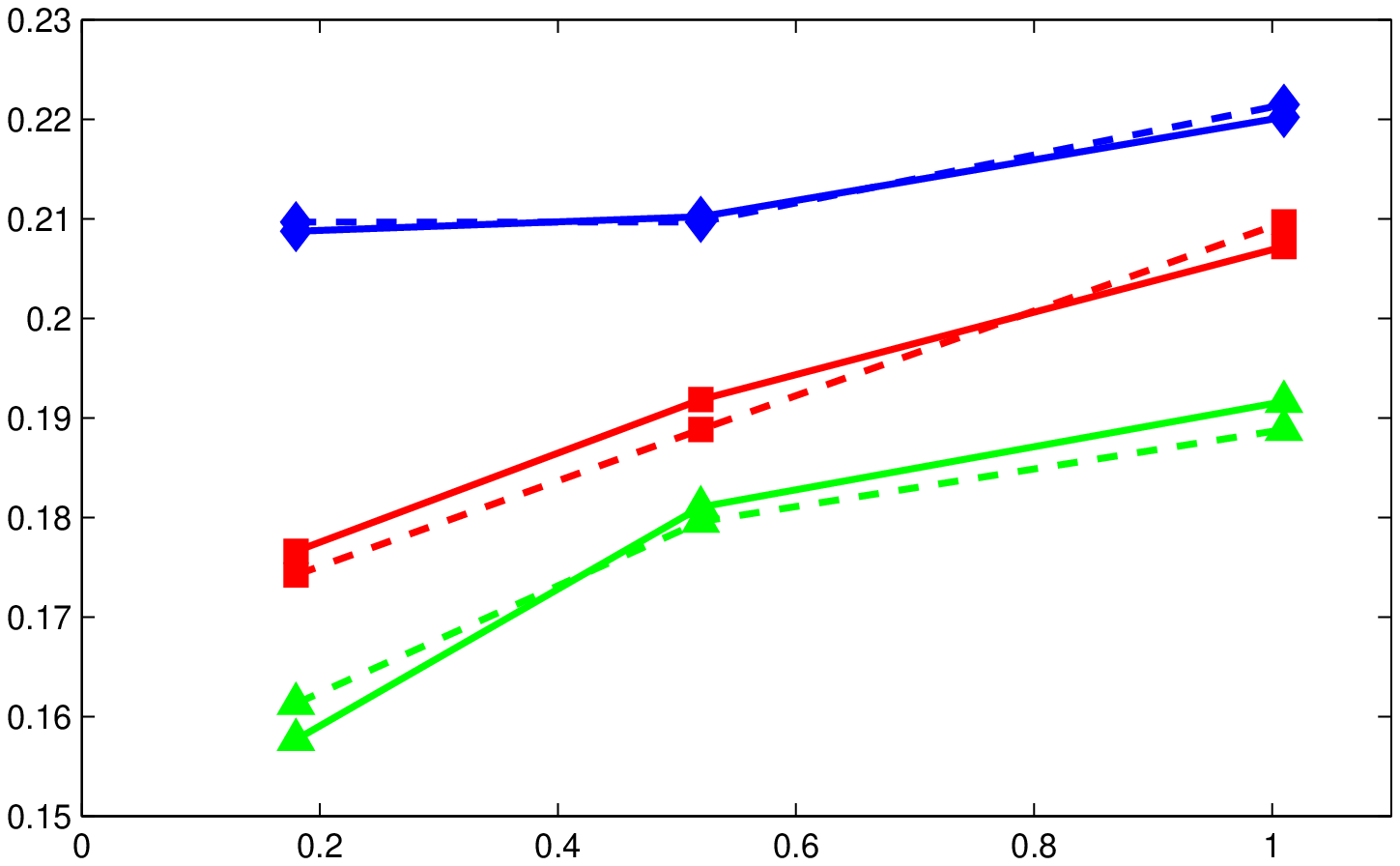}
\caption{Comparison between market option implied volatility (continuous line) and GCMG implied volatility (dashed line). Horizontal axis displays option time to maturity, vertical axis volatility level for options with moneyness equal to 1.06 (diamonds), 1 (squares), 0.94 (triangles). Left: $L=1000$, $\epsilon=-0.01$, right: $L=10000$, $\epsilon=-0.01$.}
\label{figure_options_GMG_3}
\end{figure}

\begin{figure}[h]
\centering
\includegraphics[width=0.5\textwidth]{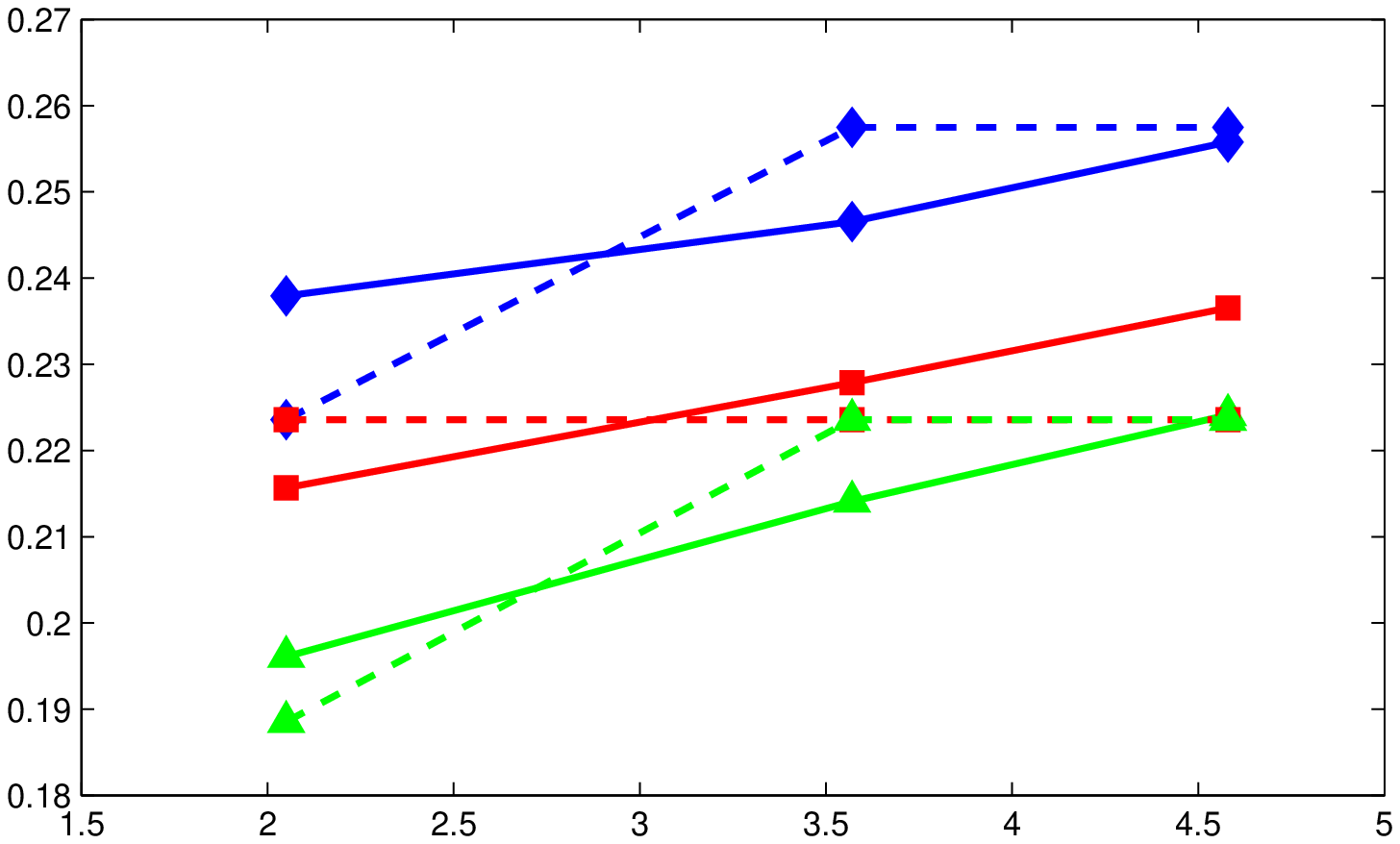}\includegraphics[width=0.5\textwidth]{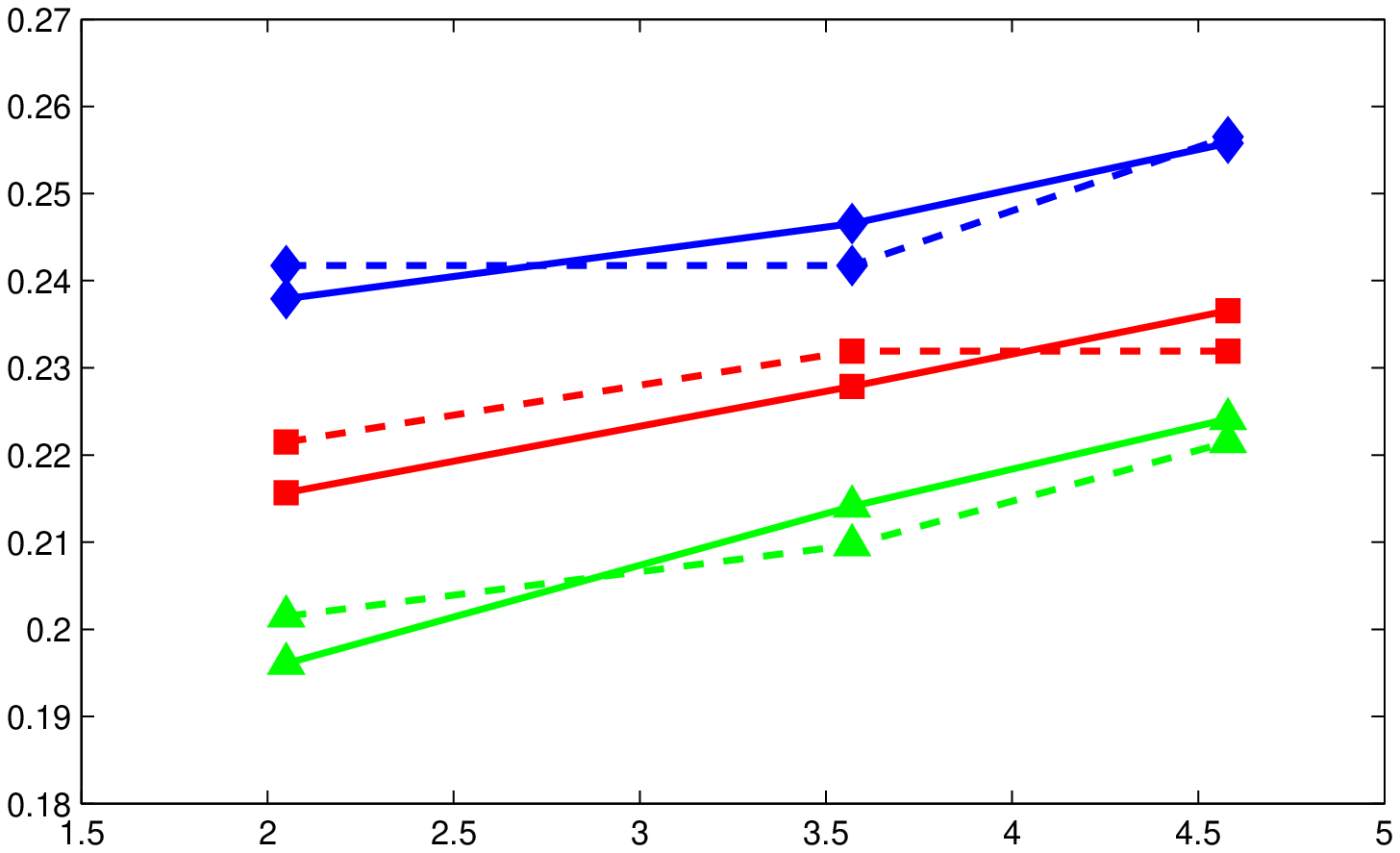}
\caption{Comparison between market option implied volatility (continuous line) and GCMG implied volatility (dashed line). Horizontal axis displays option time to maturity, vertical axis volatility level for options with moneyness equal to 1.1 (diamonds), 0.98 (squares), 0.88 (triangles). Left: $L=1000$, $\epsilon=-0.01$, right: $L=10000$, $\epsilon=-0.01$.}
\label{figure_options_GMG_4}
\end{figure}

We choose $18$ call options, with different maturities and different moneyness, and calibrate the games over these options minimizing the sum of squared difference of market implied volatility and game implied volatility. 
Results of the calibration are displayed in figures \ref{figure_options_MG_6} - \ref{figure_options_GMG_4}.

From the analysis of the calibration results (see figure \ref{figure_alpha_ratios_MG}) we can see that for both the MG and the GCMG, parameters $\alpha$ and $\alpha_{n_s}$ are greater than the critical points  and close to them for all the game parameters $M$, $L$ and $\epsilon$ used for the simulations:  this means that the option market operates in the asymmetric phase of both the MG and the GCMG in vicinity of the critical points.

About the quality of the calibration it is also useful to look at the ability of the game to fit the sample of listed options: figures \ref{figure_options_MG_6} - \ref{figure_options_GMG_4} show a good fitting at different game size and an improvement of the quality of the calibration as $M$ and $L$ increase. As one would expect a game of bigger size is able to replicate more accurately the real option market than a smaller size game. 

\begin{figure}[h]
\centering
\includegraphics[width=0.5\textwidth,height=0.25\textwidth]{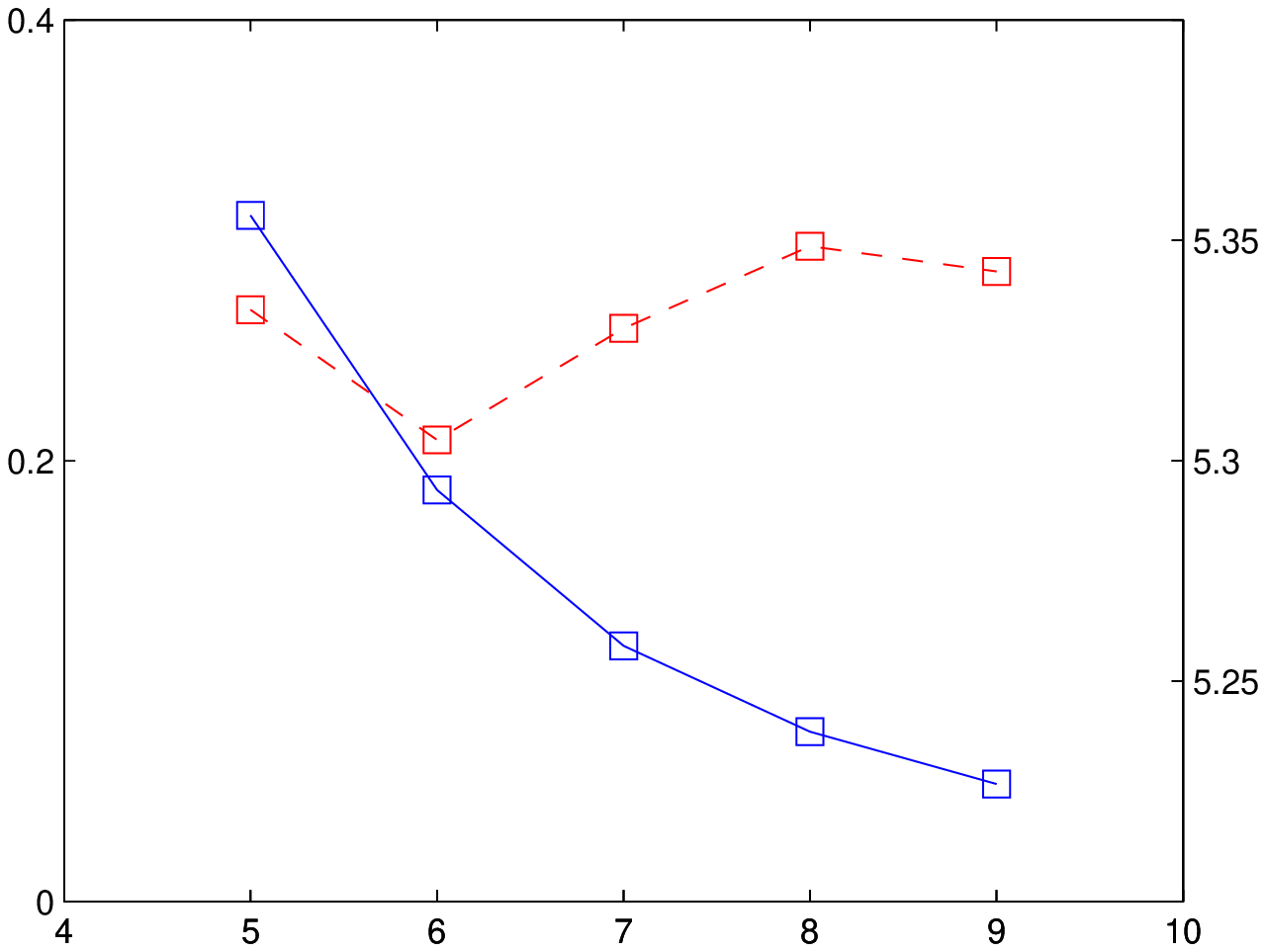}\includegraphics[width=0.5\textwidth]{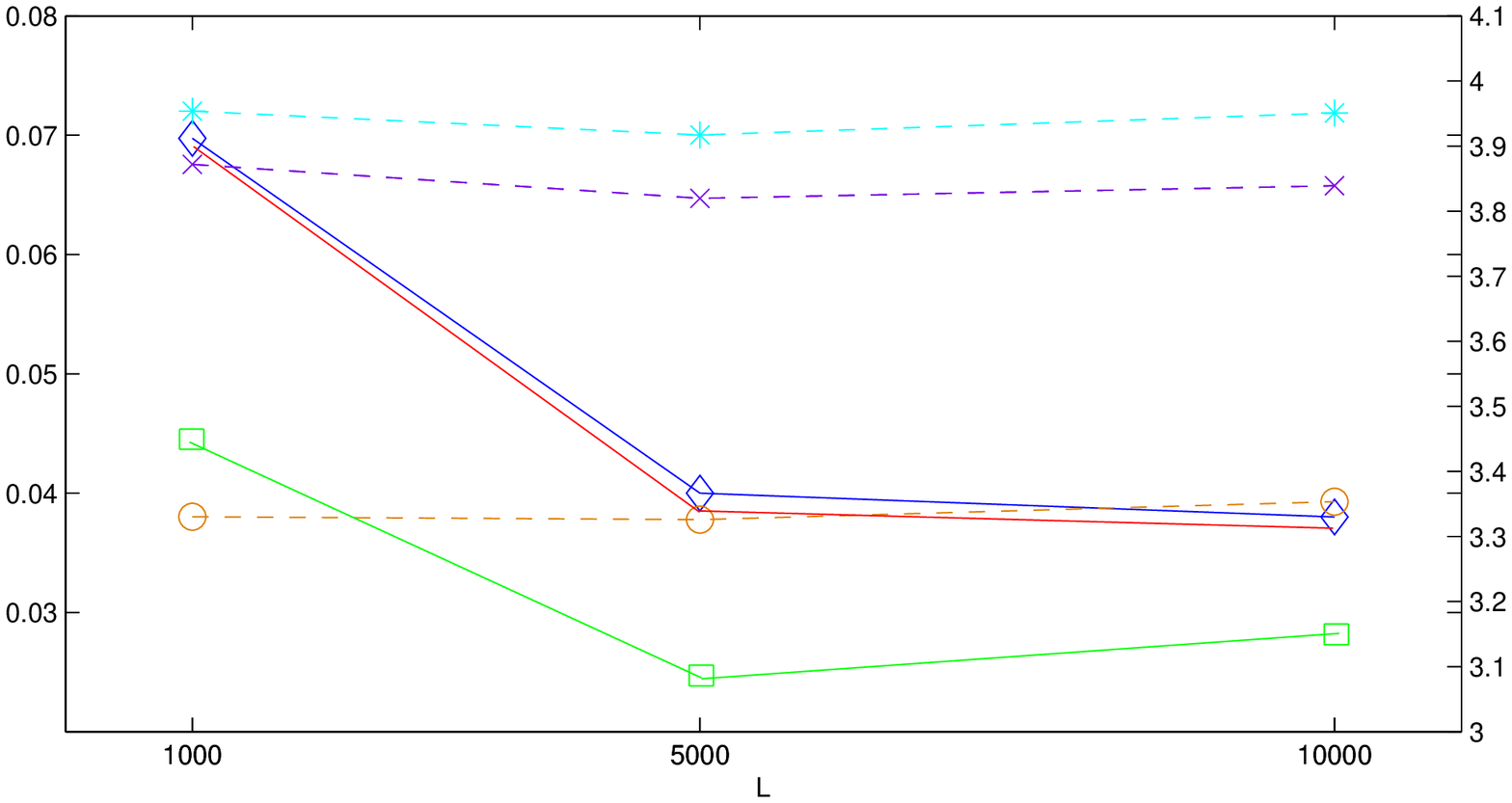}
\caption{Left: on left axis calibrated $w$ for MG (continuous line), on the right axis calibrated $w$ without time rescaling for MG (dotted line). Right: on left axis calibrated $w$ for GCMG (continuous line), on the right axis calibrated $w$ without time rescaling for MG (dotted line); from top to bottom $\epsilon=0.01$, $\epsilon=-0.01$, $\epsilon=0.1$.}
\label{figure_w}
\end{figure}

Last observation involves parameter $w$. A common feature of both games is that as $M$ and $L$ increase the inverse liquidity parameter $\bar{w}$ decreases: when the market efficiency improves, and this is what happens as $M$ and $L$ increase since in both games the minimum of $\sigma_N^2$ decreases, each agent decreases the weight of its buy or sell decision: players acting in less efficient markets characterized by scarce liquidity are required to increase the weight of the bets, like it happens in real market.
Whilst the behavior of $w$ as inverse liquidity parameter is coherent with what happens in real markets, the very low values of the calibrated $w$ deserve further investigations. Looking at the expression of $w$ coming from the implied volatility formula (\ref{mkt_equation}) and keeping in mind that the diffusion term of equations (\ref{stoch_eq1.1}) and (\ref{stoch_eq1.1_GCMG}) is linked to the characteristic time of the game dynamics (in our case the rescaled continuous time is $\tau=\frac{t}{P}$), we can conclude that a different time rescaling has the effect of increasing the level of $w$ obtained from calibration. Using a rescaled time $\tau=\frac{t}{N}$, $w$ would increase of a factor $\frac{1}{\sqrt{\alpha_c}}$. A game without time rescaling ($\tau=t$) would increase the value of calibrated $w$ of a factor $\sqrt{N}$, obtaining values of $w$ greater than 1 and more stable as $M$ and $L$ increase (see figure \ref{figure_w}).

\clearpage


\clearpage
\begin{thebibliography}{99}


\bibitem{Bouchaud}{Bouchaud J.P. and Potters M.} (2000) {\em Theory of financial risks: from statistical physics to risk management}. Cambridge University Press, Cambridge.

\bibitem{cavagna}{Cavagna A., Garrahan J~P., Giardina I. and Sherrington D.} (1999) A thermal model for adaptive competition in a market. {\em Phys. Rev. Lett.}, {\bf 83}, 4429.

\bibitem{challet_quan}{Challet D., Chessa A., Marsili M. and Zhang Y.~C.} (2001) From Minority Games to real markets. {\em Quantitative Finance}, {\bf 1}, 168--176.

\bibitem{challet}{Challet D. and Zhang Y.~C.} (1997) Emergence of cooperation and organization in an evolutionary game. {\em Physica A}, {\bf 246}, 407.

\bibitem{marsili_GCMG}{Challet D., Marsili M.} (2003) Criticality and market efficiency in a simple realistic model of the stock market. {\em Phys. Rev. E}, {\bf 68}, 036132, 4 pages.

\bibitem{challet1}{Challet D., Marsili M. and Zhang Y.~C.} (2001) Stylized facts of financial markets and market crashes in Minoriy Games {\em Physica A}, {\bf 294}, 514--524.

\bibitem{coolenlibro}{Coolen A.C.C.} (2005) {\em The Mathematical Theory of Minority Games - Statistical Mechanics of Interacting Agents}. Oxford University Press.

\bibitem{coolen10}{Coolen A.C.C. and J.A.F. Heimel} (2001) Dynamical solution of the on-line minority game. {\em J. Phys. A}, {\bf 34}, 10783--10804.

\bibitem{coolen}{Coolen A.C.C.} (2005) Generating functional analysis of minority games with real market histories. {\em J. Phys. A}, {\bf 38}, 2311--2347.

\bibitem{marsililibro}{Challet D., Marsili M. and Zhang Y.~C.} (2004) {\em Minority Games: Interacting agents in financial markets}. Oxford Finance Series.


\bibitem{giardina}{Giardina I., Bouchaud J.P. and Mezard M.} (2001) { Microscopic models for long ranged volatility correlations}. {\em Physica A}, {\bf 299}, 28-39.

\bibitem{Jef}{Jefferies P., Hart M.L., Hui P.M. and Johnson N.F.} (2001) {From market games to real-world markets}. {\em Eur.Phys. J. B} {\bf 20}, 493-501.

\bibitem{john}{Johnson N.F., Hart M., Hui P.M. and Zheng D} (2000) {Trader dynamics in a model market}. {\em J. Theo. App. Fin.}, {\bf 3}, 443.

\bibitem{kar}{Karatzas I. and Shreve S.E.} (1991) {\em Brownian Motion and Stochastic Calculus}. Springer-Verlag, New York.


\bibitem{lux}{Lux T. and Marchesi M.} (1999) {Scaling and criticality in a stochastic multi-agent model of financial market}. {\em Nature}, {\bf 397}, 498-500.

\bibitem{marsili1}{Marsili M. and Challet D.} (2001) Continuum time limit and stationary states of the minority game. {\em Phys. Rev. E}, {\bf 64}, 056138, 12 pages.


\bibitem{nov}{Novikov A.A.} (1971) On moment inequalities for stochastic integrals. {\em Theory Prob. Appl.}, {\bf 17}, 717-720.

\bibitem{ortisi}{Ortisi M.} (2008) Polynomial-rate convergence to the stationary state for the continuum-time limit of the Minority Game. {\em Journal of Applied Probability}, {\bf 45}, 376-387. 

\bibitem{shreve}{Shreve S.E.} (2004) {\em Stochastic Calculus for Finance, V. II}. Springer Finance, NY. 

\bibitem{stan}{Stanley H.S. and Mantegna R.N.} (2000) {\em An introduction to econophysics: correlations and complexity in finance}. Cambridge University Press, Cambridge.


\end{thebibliography}
\end{document}